\documentclass[11pt]{article}
\usepackage[a4paper,hmarginratio=1:1,vmarginratio=2:3,totalwidth=15.6cm,
totalheight=21.5cm]{geometry}
\usepackage{bm,epstopdf,epsfig,amsmath,amssymb,amsfonts,
latexsym,comment,cancel,verbatim}
\usepackage[font=md,captionskip=8pt]{subfig}
\usepackage{amsmath}
\usepackage{amssymb}
\usepackage{amsfonts}
\usepackage{epsfig}

\renewcommand{\baselinestretch}{1.25}

\newcommand{\cD}{{\cal D}}

\newcommand{\cL}{{\cal L}}

\newcommand{\cO}{{\cal O}}

\newcommand{\cZ}{{\cal Z}}

\newcommand{\opsi}{{\overline\psi}}

\newcommand{\ra}{\rightarrow}
\newcommand{\be}{\begin{equation}}
\newcommand{\ee}{\end{equation}}
\newcommand{\bea}{\begin{eqnarray}}
\newcommand{\eea}{\end{eqnarray}}

\long\def\symbolfootnote[#1]#2{\begingroup
\def\thefootnote{\fnsymbol{footnote}}\footnote[#1]{#2}\endgroup}
\begin{document}
\begin{flushright}
CERN-PH-TH/2010-118\\
CPHT-RR037.0510\\
LPT-ORSAY 10-32\\
NSF-KITP-10-057
\end{flushright}

\thispagestyle{empty}
\vspace{0.1cm}

\begin{center}
{\Large {\bf  Non-linear MSSM}}
\vspace{1.cm}
\end{center}

\begin{center}
{\bf I. Antoniadis$^{\,a, b\,}$,
E. Dudas$^{\,b, c\,}$,
D.~M. Ghilencea$^{\,a,\,d\,}$, 
P. Tziveloglou$^{\,a,\,e,\,}$\footnote{
E-mail addresses:\,\, Ignatios.Antoniadis@cern.ch,
Emilian.Dudas@cpht.polytechnique.fr,

$\,\,\,\,$Dumitru.Ghilencea@cern.ch,
pt88@cornell.edu}}\\

 \vspace{0.5cm}
 {\small $^a\,$Theory Division, CERN, 1211 Geneva 23,
 Switzerland.}\\[2pt]
 {\small $^b\,$Centre de Physique Th\'eorique, Ecole Polytechnique, CNRS, 91128
   Palaiseau, France.}\\[2pt]
 {\small $^c\,$LPT, UMR du CNRS 8627, B\^at 210, Universit\'e de Paris-Sud,
 91405 Orsay Cedex, France.}
 \\[2pt]
{\small $^d\,$Theoretical Physics Department,
 IFIN-HH Bucharest, PO Box MG-6 Bucharest, Romania.}
\\[2pt]
{\small $^e\,$Department of Physics, Cornell University, Ithaca, NY 14853 USA.}
 \end{center}

\vspace{1cm}
\def\baselinestretch{1.1}
\begin{abstract}
\noindent
Using the formalism of constrained superfields, we derive the most
general effective action of a light goldstino coupled to the minimal
supersymmetric standard model (MSSM) and study its phenomenological
consequences. The goldstino-induced couplings
become important when the (hidden sector)
scale of spontaneous supersymmetry breaking, $\sqrt f$, is relatively
low, of the order of few TeV. In particular, we compute the Higgs
potential and show that the (tree level)
mass of the lightest Higgs scalar can be increased
to the LEP bound for $\sqrt f\sim 2$ TeV to 7~TeV.
Moreover, the effective quartic Higgs coupling
is increased due to additional tree-level contributions proportional to the
ratio of visible to hidden sector supersymmetry breaking scales.
This increase can alleviate the amount of fine tuning of the electroweak scale
that exists in the MSSM.
Among the new goldstino couplings, beyond those in
MSSM, the most important ones generate an invisible decay of the
Higgs boson into a goldstino and neutralino (if $m_h>m_{\chi_1^0}$), 
with a partial decay rate that  can  be comparable to the SM 
channel $h^0\ra \gamma\gamma$. A similar decay of $Z$ boson is
possible if $m_Z>m_{\chi_1^0}$ and brings a lower bound on
$\sqrt f$ that must be of about $700$ GeV.
Additional decay modes  of the Higgs or Z bosons
into a pair of light goldstinos, while possible,  are suppressed by an
extra $1/f$ factor and have  no significant impact on the model.
\end{abstract}

\newpage

\section{Introduction}

 Spontaneous supersymmetry breaking at low energies predicts a nearly
 massless goldstino. More precisely, it plays the role of the
 longitudinal component of the gravitino, which acquires a Planck
 suppressed mass $f/M_{Planck}$, in the milli-eV range if the supersymmetry
 breaking scale $\sqrt f$ is in the multi-TeV region. By the
 equivalence theorem \cite{cddfg},
 it interacts with a strength $1/\sqrt f$ which
 is much stronger than the Planck suppressed couplings of the
 transverse gravitino, and is therefore very well described by the
 gravity-decoupled limit of a massless Goldstone fermion. An example
 of such a situation is provided by gauge mediation where, however, the
 typical scale of supersymmetry breaking is expected to be a few
 orders of magnitude higher than the soft breaking terms of the
 Standard Model (SM) superparticles, due to their double suppression by
the loop factor and by the messengers mass.

In this work, we perform a model independent analysis of the low
energy consequences of a light goldstino by treating $\sqrt f$ as a
free parameter, that can be as low as a few times the scale of soft
breaking terms which we denote generically $m_{soft}$. Furthermore, we
will assume that all extra states (that may exist
beyond those of the MSSM) are
heavier than $\sqrt f$. In such a framework, there are two generic
energy regimes that can be studied:
{\bf (i):} at TeV energies, comparable
(or higher) than $m_{soft}$, one has the usual MSSM together with a
goldstino; 
{\bf (ii):} at low energies, lower than all sparticle masses, one is left
just with a goldstino coupled to the SM fields. In both cases, the
goldstino effective interactions can be determined by non-linear
supersymmetry. In the first case, it couples to ordinary
supermultiplets of linear supersymmetry, while in the second
case the superparticles have been integrated out.

The self-interactions of the goldstino are given by the famous
Volkov-Akulov action~\cite{Volkov:1973ix}. Their geometric method
gives also a universal coupling to matter through its energy momentum
tensor, of the form $(1/f^2) T_{\mu\nu}t^{\mu\nu}$, where
$T_{\mu\nu}$, $t_{\mu\nu}$ are the stress tensors of matter and of
(free) goldstino, respectively~\cite{Clark:1996aw, Clark:1998aa}. It
was realized however that this coupling is not the most general
invariant under non-linear supersymmetry~\cite{Clark:1998aa,Brignole:1997pe,
  Luty:1998np}. General invariant couplings can be
derived using two different superfield formulations. One of them
promotes any ordinary field to a superfield by introducing a modified
superspace that takes into account the non-linear supersymmetry
transformations of the goldstino~\cite{Ivanov:1977my,Samuel:1982uh,
  Antoniadis:2004uk}. The other uses the formalism of constrained
superfields: these are usual superfields, but  are subject to
constraints
that eliminate the superpartners in terms of the light degrees
of freedom and the
 goldstino~\cite{Rocek:1978nb,Lindstrom:1979kq,Casalbuoni:1988xh,SK}.

In this work, we use the method of constrained superfields in order to
determine the general couplings of the goldstino to MSSM superfields,
focusing in
the first energy region mentioned above $E\sim m_{soft}<\sqrt{f}$. The only
constrained superfield is then that of the goldstino $X_{nl}$,
satisfying the constraint $X_{nl}^2=0$, which couples to MSSM
superfields via the corresponding soft terms: the recipe is to replace
the spurion $S\equiv m_{soft}\theta^2$ by $(m_{soft}/f)X_{nl}$ and
solve for its F-auxiliary component, as usual, in order to determine
all effective
interactions that can be expanded in inverse power series of $f$.
In the second energy region, lower than $m_{soft}$, the superpartners
can be integrated out (this can be done by additional constraints
 on the MSSM superfields \cite{SK}),
and one has the goldstino coupled to SM fields only.
In this case, it was found that the dominant effective operators
are of dimension-six~\cite{Antoniadis:2004uk} and can induce an
important invisible decay width of the  Higgs boson, if the goldstino
carries lepton number~\cite{Antoniadis:2004se}.
For a related effective approach to these problems,
goldstino couplings and applications see 
\cite{Brignole:2003cm,AlvarezGaume:2010rt}.

Obviously, the goldstino couplings to MSSM become important if the
supersymmetry (SUSY) breaking scale is low. On the other hand, validity of
the effective Lagrangian requires that $f$
be higher than the soft breaking terms, so that
 $m_{soft}^2/f$ is a good expansion parameter.
It turns out that the most important effects
of these couplings are in the Higgs sector. In particular, the quartic
Higgs coupling is increased by a term proportional to the ratio of
visible to hidden sector SUSY breaking, with two important consequences:
(i)~it can increase  the tree-level value of the lightest Higgs mass that can
then reach and cross the LEP bound\footnote{For other possibilities
to increase the Higgs mass in effective models see
\cite{Brignole:2003cm,Antoniadis:2009rn,Dine:2007xi,Cassel:2009ps,
Carena:2009gx}.} of 114.4 GeV \cite{Barate:2003sz};
(ii) it can alleviate  the 
fine tuning of the electroweak scale in MSSM due to the
relatively high experimental bounds on $m_{soft}$
and large quantum corrections usually
 required in MSSM to satisfy the LEP bound.
Additional effects that we investigate relate to the goldstino-induced
couplings in the MSSM Lagrangian, upon integration of the sgoldstino.
All couplings goldstino - MSSM fields are computed
and these can be used for phenomenological studies.
As an example we show that for a light neutralino, the SM-like Higgs can decay
into a goldstino (which is the lightest supersymmetric particle (LSP))
and the lightest neutralino (next to LSP (NLSP) in this case), with
a decay rate that can be comparable to the SM
partial decay  $h^0\ra \gamma\gamma$. A similar decay of $Z$ 
is possible, which provides a lower bound on $\sqrt f\approx 700$~GeV. 
Other decays of the Higgs and Z bosons into pairs of goldstinos are
 possible, but they have additional $(1/f)$ suppression, with
little impact on the allowed parameter space.

The paper is organized as follows. Section~\ref{section1} discusses
general issues in non-linear realizations of supersymmetry and
goldstino couplings. Section~\ref{section2} presents
the ``non-linear'' MSSM model obtained by the general coupling of 
the MSSM to goldstino, using the method described above. 
Section~\ref{section3} presents the
new couplings of the model, not present in MSSM, some of which are dimension-four
in fields,  suppressed by up to the second power of $1/f$.
Section~\ref{section4} analyzes the implications for the Higgs
masses. Section~\ref{section5}  presents other phenomenological 
consequences, such as the implications for the  fine-tuning of the 
electroweak scale, some interesting limits, and the
invisible decays of the Higgs and Z bosons 
 together with their constraints.
Finally, Section~\ref{conclusions} contains  our concluding remarks.

\section{Non-linear realizations 
and constrained goldstino superfield}\label{section1}

An important role in constructing a non-linear supersymmetric version of the
MSSM is played by the goldstino chiral superfield (SM gauge singlet) $X_{nl}$.
One can use the component fields formalism
to describe the  corresponding  Volkov-Akulov action. However,
one can use the more convenient superfield formalism, 
endowed with constraints;
for the goldstino superfield this constraint
is $X_{nl}^2=0$ \cite{Rocek:1978nb,Lindstrom:1979kq,Casalbuoni:1988xh,SK}.
One can start with the Lagrangian
\bea\label{X}
\cL_X=\!
\int\! d^4\theta \,X^\dagger X
+\bigg\{\!\int\! d^2\theta \,f\,X+h.c.\bigg\}
\!=\!\vert \partial_\mu\phi_X\vert^2\!+\!F_X^\dagger F_X\!+\!
\Big[\frac{i}{2}\overline\psi_{X}\overline\sigma^\mu
\partial_\mu\psi_{X}+f\,F_X+\!h.c.\!\Big]
\eea
\smallskip\noindent
with the aforementioned constraint. This constraint is solved by
\bea\label{goldstino1}
X_{nl}&=&\phi_X +\sqrt 2\,\, \theta\psi_X
+\theta\theta\,\,F_X,\qquad\rm{with}\qquad
\phi_X=\frac{\psi_X \psi_X}{2\, F_X}
\eea
which, when used in eq.(\ref{X})  recovers \cite{SK}
 the Volkov-Akulov Lagrangian. After using the equations of motion
$F_X=-f+....$ where $f$ (that can be chosen real)
is the hidden sector SUSY breaking
scale. Therefore, in the infrared description of the SUSY breaking
(which is model independent), the scalar component
(sgoldstino) becomes a function of the goldstino.

To find the goldstino couplings to matter fields
(discussed for the MSSM in the next section), 
consider first a  supersymmetric theory with chiral 
multiplets $\Phi_i\equiv(\phi_i,\psi_i, F_i)$ and vector multiplets
$V\equiv (A_\mu^a, \lambda^a, D^a)$ coupled in a general way 
to $X_{nl}$:
%\medskip
\bea
{\cal L}& =&\!\! \int d^4 \theta\,\, \Big[ X_{nl}^{\dagger} X_{nl} 
+ \Phi_i^{\dagger} (e^{V} \Phi)_i - ({m_i^2}/{f^2})\,
 X_{nl}^{\dagger} X_{nl} \Phi_i^{\dagger} (e^{V} \Phi)_i  \Big]
+\bigg\{ \int d^2 \theta \,\Big[ f X_{nl} + W(\Phi_i) 
\nonumber\\[-3pt]
&+&\!
\frac{B_{ij}}{2 f} X_{nl}\, \Phi_i \Phi_j
+ \frac{A_{ijk}}{6\,f} X_{nl} \Phi_i \Phi_j \Phi_k
+ \frac{1}{4} \Big(1+ \frac{2 \,m_\lambda}{f} X_{nl}\Big)\,
\mbox{Tr}\, W^{\alpha} W_{\alpha} 
\Big]
+ {\rm h.c.}\bigg\}, \quad\label{01}
\eea
where $m_i^2,B_{ij}, A_{ijk}$ are soft terms for the scalars and
$m_{\lambda}$ is the gaugino mass. From this, one can find the
Goldstino ($\psi_X$) couplings to ordinary matter/gauge superfields.
These couplings can be checked to be equivalent to those
obtained by the equivalence theorem \cite{cddfg}, 
from a theory with the corresponding explicit soft breaking,
in which the goldstino couples as:
\be
({1}/{f}) \ \partial^{\mu} \psi_X \ J_{\mu} = \
 - ({1}/{f}) \,\psi_X \ \partial^{\mu}  J_{\mu} +\mbox{(total space-time 
derivative)} , \label{02}
\ee
Here $J_{\mu}$ is the supercurrent of the theory corresponding to
that in (\ref{01}) in which the goldstino is  essentially replaced 
by the spurion, with the corresponding explicit soft breaking terms:
\medskip
\bea
{\cal L}' &=&\!\!\!\! \int\!\! d^4 \theta\,
\Big[1-m_i^2\,\theta^2\overline\theta^2\Big] 
\Phi_i^{\dagger} (e^{V} \Phi)_i 
+\!\! \int\! d^2 \theta \,\Big[ 
W(\Phi_i)\! - (1/2)\,B_{ij}\,\theta^2 \Phi_i \Phi_j
-\! (1/6)\,A_{ijk}\,\theta^2\, \Phi_i \Phi_j \Phi_k 
\nonumber\\
&\!+&\!\!\frac{1}{4}\,(1-2 m_\lambda\theta^2) \,
\mbox{Tr}\, W^{\alpha} W_{\alpha} 
\Big] + {\rm h.c.}\ , \label{01prime}
\eea
With this, eq.(\ref{02}) shows that, on-shell, all goldstino 
couplings are proportional to soft terms. 
Indeed, the supercurrent of (\ref{01prime}) is given by
(with $\cD_{\mu,ij}=\delta_{ij}\,\partial_\mu +i\,g\,A^a_\mu\,T^a_{ij}$)
%\medskip
\bea
J^\mu_\alpha=
-[\sigma^\nu \overline\sigma^\mu\psi_i]_\alpha
\,[\cD_{\nu,\,ij}\phi_j]^\dagger
+i\,[\sigma^\mu\overline\psi_i]_{\alpha}
F_i
-\frac{1}{2\sqrt 2}[\sigma^\nu\overline\sigma^\rho\,\sigma^\mu
\overline\lambda^a]_\alpha\,F_{\nu\rho}^a
+\frac{i}{\sqrt 2} D^a\,[\sigma^\mu\overline\lambda^a]_\alpha
\eea
so
\vspace{-0.2cm}
\bea\label{jjj}
\partial_{\mu} J^{\mu}_\alpha \ = \, \psi_{i,\alpha} 
\,\,(m_{i}^2  {\phi^\dagger}_j
 +  B_{ij} \phi_j + (1/2) A_{ijk} \phi_j \phi_k\, ) +
\frac{m_{\lambda}}{\sqrt{2}} 
\Big[({\sigma^{\mu\nu}})_\alpha^{\,\,\beta}\,
\lambda^a_\beta F_{\mu\nu}^a + D^a\,\lambda^a_\alpha\Big] \ .\eea
From (\ref{02}), (\ref{jjj}) one then recovers the 
couplings with one goldstino.
However, the superfield formalism in (\ref{01})  
has the  advantage that is easier to use when evaluating  couplings 
with more than one goldstino, by simply writing all effective 
operators (involving $X_{nl}$) to a fixed order in $1/f$~\cite{SK}.
It is more difficult to find these 
from (\ref{01prime}) and in Section~\ref{section2} we use the former 
method.

 Finally, in addition to usual SUSY and goldstino couplings
eq.(\ref{01}) 
also brings new goldstino-independent couplings induced by 
eliminating $F_X$.
Indeed, from (\ref{01})
\be
\Big(1-\frac{m_i^2}{f^2}\,\vert \phi_i\vert^2\Big)\,
F_X^\dagger=-\Big(  f +\frac{B_{ij}}{2\,f}\, \phi_i \phi_j + 
 \frac{A_{ijk}}{6\,f}\, \phi_i \phi_j \phi_k +
 \frac{m_{\lambda}}{2\,f} \lambda \lambda+\cdots\Big), \
\ee
%\medskip\noindent
So $\vert F_X\vert^2$ generates 
new couplings in  onshell $\cL$, such as quartic scalar
 terms. When applied to MSSM, this brings in particular new 
corrections  to the Higgs scalar potential (see later).

\section{The ``non-linear'' MSSM.}\label{section2}

We apply the above method to couple the
constrained superfield $X_{nl}$ to the
SUSY part of the MSSM, to find
the ``non-linear'' supersymmetry
version of MSSM \cite{SK}. We stress that at energy scales below $m_{soft}$,
similar constraints can be applied to the MSSM superfields themselves,
corresponding to integrating out the corresponding superpartners.
Here, the only difference from the
ordinary MSSM is in the supersymmetry breaking sector.
Supersymmetry is broken spontaneously via a
 vacuum expectation value (VEV) of $F_X$, fixed
by its equation of motion (see later). 
The  Lagrangian of the ``non-linear MSSM'' model is \cite{SK},
\bea
\label{LL}
\cL=\cL_0+\cL_X+\cL_{H}+\cL_m+\cL_{AB}+\cL_{g}
\eea
%\medskip\noindent
Let us detail these terms. 
$\cL_0$ is the usual MSSM SUSY Lagrangian, in standard notation:
\smallskip
\bea\label{mssmsusy}
\cL_0 &=&\!\!\sum_{\Phi, H_{1,2}} \int d^4\theta \,\,
\Phi^\dagger\,e^{V_i}\,\Phi+
\bigg\{\int
d^2\theta\,\Big[\,\mu\,H_1\,H_2+
H_2\,Q\,U^c+Q\,D^c\,H_1+L\,E^c\,H_1\Big]+h.c.\bigg\}
\nonumber\\[-6pt]
&&
\hspace{1cm}+\sum_{{\rm SM\,groups}}
\frac{1}{16\, g^2\,\kappa}
\int d^2\theta
\,\mbox{Tr}\,[\,W^\alpha\,W_\alpha] +h.c.,
\qquad\quad \Phi:Q,D^c,U^c,E^c,L\, ,
\eea
$\kappa$ is a constant canceling the trace factor and the
gauge coupling $g$ is shown explicitly.

The SUSY breaking couplings originate from the MSSM fields
couplings to the goldstino superfield; this is done by the
replacement $S\!\ra\! (1/f)\,m_{soft} X_{nl}\,\,$~\cite{SK},
 where $S$ is the spurion,
with $S=\theta\theta \,m_{soft}$ and $m_{soft}$ is a generic
notation for the soft terms (denoted below $m_{1,2}, B,
m_0$).
One has for the Higgs sector
\medskip
\bea
\cL_{H}\!\! &=&\!\!\sum_{i=1,2} c_i
\int d^4\theta \,\,X_{nl}^\dagger X_{nl}\,\,
H_i^\dagger\,e^{V_i}\,H_i
\nonumber\\[-7pt]
&=&\!\!\! \sum_{i=1,2} c_i \,\Big\{
\vert\phi_X\vert^2\,\Big[
\vert \cD_\mu\, h_i\vert^2
+F_{h_i}^\dagger F_{h_i}+h_i^\dagger\,\frac{D_i}{2}\,h_i+
\Big(\frac{i}{2}\overline\psi_{h_i}\overline\sigma^\mu\,\cD_\mu\psi_{h_i}
-\frac{1}{\sqrt 2} \,h_i^\dagger \lambda_i\,\psi_{h_i}+h.c.\Big)\Big]
\nonumber\\[-6pt]
&+&\!\!\!
\frac{1}{2}\, h_i^\dagger\,(\cD_\mu+\overleftarrow
\cD_\mu)\,h_i\,\,\partial^\mu\vert \phi_X\vert^2
+\overline\psi_X\overline\psi_{h_i}\,\psi_X\psi_{h_i}
-\frac{1}{2}\,[\phi_X^\dagger\,(\partial^\mu-\overleftarrow
\partial^\mu)\,\phi_X]\,[ h_i^\dagger (\cD_\mu - \overleftarrow
\cD_\mu)\,h_i]\nonumber\\[-2pt]
&+&\!\!\!
\Big[
-\frac{i}{2}\phi_X^\dagger \psi_X\, \sigma^\mu\,\overline\psi_{h_i}
(\cD_\mu-\overleftarrow \cD_\mu) {h_i}
-\frac{1}{\sqrt 2}\, \phi_X^\dagger \psi_X \,\,
h_i^\dagger\lambda_i\,{h_i}
-\phi_X^\dagger \psi_X \,F_{h_i}^\dagger\psi_{h_i}
+\phi_X^\dagger F_X\,F_{h_i}^\dagger {h_i}
\nonumber\\[-2pt]
&+&\!\!\!
\frac{i}{2}
\,(\overline\psi_X\,\overline\sigma^\mu\,\psi_X)\,(h_i^\dagger
\,\cD_\mu\,{h_i})
+\frac{i}{2}\,(\phi_X^\dagger \partial_\mu\,\phi_X)
\,(\overline\psi_{h_i}\,\overline\sigma^\mu\,\psi_{h_i})
+\frac{i}{2}
\overline\psi_X\,\overline\sigma^\mu\,
(\partial_\mu-\overleftarrow\partial_\mu)\,\phi_X\,\,
(h_i^\dagger \psi_{h_i})
\nonumber\\[-1pt]
&-&
\overline\psi_X\,F_X\,\,\overline\psi_{h_i}\,{h_i}+h.c.\Big]
+
\Big[
\partial_\mu\phi_X^\dagger \partial^\mu\phi_X+F_X^\dagger F_X
+\Big(\frac{i}{2}
\overline\psi_X\,\overline\sigma^\mu\partial_\mu\psi_X+h.c.\Big)\Big]
\vert\,{h_i}\vert^2\Big\},
\eea

\medskip\noindent
Here  $\cD, \partial$,
($\overleftarrow\cD,\overleftarrow\partial$) act only on the first
field to their right (left) respectively and $h_i$, $\psi_{h_i}$,
$F_{h_i}$ denote SU(2) doublets. Also
\bea
c_{1}=-{m_1^2}/{f^2},\qquad c_2=-{m_2^2}/{f^2}\, .
\eea

\medskip\noindent
Similar terms exist for all matter fields $Q, U^c, D^c,  L, E^c$:
\medskip
\bea
\cL_m=\sum_{\Phi} c_\Phi\int d^4\theta \,\,
X_{nl}^\dagger X_{nl}\,\Phi^\dagger
e^V\,\Phi,\qquad c_\Phi
=-\frac{m_\Phi^2}{f^2},\quad \Phi: Q, U^c, D^c, L, E^c,
\eea
%%%
One can eventually set $m_\Phi=m_0$ (all $\Phi$).
The\,bi- and trilinear
SUSY breaking couplings are
\medskip
\bea
\cL_{AB}
\!\!&=&\!\!\frac{B}{f}\,\int d^2\theta \,X_{nl}\,H_1\,H_2
\nonumber\\[-2pt]
&+&
\frac{A_u}{f}\int d^2\theta\,X_{nl}\,H_2\,Q\,U^c+
\frac{A_d}{f}\,\int d^2\theta\,X_{nl}\,Q\,D^c\,H_1+
\frac{A_e}{f}\,\int d^2\theta\,X_{nl}\,L\,E^c\,H_1+h.c.
\nonumber\\[-3pt]
&=&\frac{B}{f}\,\Big\{
\phi_X\,\Big[\,h_1.F_{h_2}+F_{h_1}.h_2-\psi_{h_1}.\psi_{h_2}\Big]
-h_1.(\psi_{X}\psi_{h_2})-(\psi_X\psi_{h_1}).h_2+F_X\,h_1.h_2
\Big\}
\nonumber\\[-3pt]
&+&\Big\{
\frac{A_u}{f}\Big[
\phi_X\,h_2.(\phi_Q\,F_U\!-\!\psi_Q\,\psi_U+\! F_Q\,\phi_U)
-\phi_X\,(\psi_{h_2}.\phi_Q \psi_U+\psi_{h_2}.\psi_Q \phi_U
-F_{h_2}.\phi_Q\,\phi_U)
\nonumber\\[-3pt]
&-&
\psi_X\,(h_2.\phi_Q\,\psi_U+h_2.\psi_Q\,\phi_U+\psi_{h_2}.\phi_Q\,\phi_U)
+F_X\,h_2.\phi_Q\,\phi_U\Big]
%\nonumber\\[-2pt]
-\Big[U\ra D, H_2\ra H_1\Big]
\nonumber\\
&-&\Big[U\ra E, H_2\ra H_1,Q\ra L\Big]\Big\}
+h.c.
\eea

\medskip\noindent
where $B\equiv B_0\,m_0\mu$.
For simplicity, Yukawa matrices are not displayed; to recover them
just replace above and in formulae below any pair of
fields $\phi_Q \phi_U\ra \phi_Q\gamma_u \phi_U$,
$\phi_Q\phi_D\ra \phi_Q\gamma_d \phi_D$,
$\phi_L\,\phi_E\ra \phi_L\gamma_e \phi_E$;
similar for the fermions and auxiliary fields,
 with $\gamma_{u,d,e}$ $3\times 3$ matrices.

Finally, the supersymmetry breaking
couplings in the gauge sector are
\medskip
\bea\label{gb}
\cL_{g}\!\!&=&
\sum_{i=1}^3
\frac{1}{16\, g^2_i\,\kappa}
\frac{2\,m_{\lambda_i}}{f}
\int d^2\theta
\,X_{nl}\,\mbox{Tr}\,[\,W^\alpha\,W_\alpha]_i +h.c.
\nonumber\\[-4pt]
&=&\sum_{i=1}^3\frac{m_{\lambda_i}}{2\,f}\,
\Big\{\phi_X\,\,\Big[
2\,i\,\lambda^a\,\sigma^\mu\,\Delta_\mu\,\overline\lambda^a
-\frac{1}{2}\,F^{a\,\mu\nu}F_{\mu\nu}^a
+D^a D^a-\frac{i}{4}\epsilon^{\mu\nu\rho\sigma}
\,F_{\mu\nu}^a\,F_{\rho\sigma}^a \Big]
\nonumber\\[-6pt]
&&\qquad\qquad\quad -\,\sqrt 2 \psi_X\,\sigma^{\mu\nu}\lambda^a\,F_{\mu\nu}^a
-\sqrt 2\,\psi_X\,\lambda^a\,D^a
+F_X\,\lambda^a \lambda^a\Big\}_i
+h.c.
\eea

\medskip\noindent
with $m_{\lambda_{1,2,3}}$ the masses of the three gauginos
and gauge group index $i$ for $U(1)$, $SU(2)$, $SU(3)$ respectively.
Above we introduced the notation
$\Delta_\mu\overline\lambda^a=\partial_\mu\overline\lambda^a
-g\,t^{abc}\,V_\mu^b\,\overline\lambda^c$.
Equations~(\ref{X}) to (\ref{gb}) define the model, with spontaneous
supersymmetry breaking ensured by non-zero $\langle F_X\rangle$.

Since $\phi_X\sim 1/f$,  the Lagrangian contains terms of order
higher than $1/f^2$.  In the calculation of the onshell Lagrangian
we shall restrict the calculations to up to and including $1/f^2$ terms.
This requires solving for $F_\phi$ of matter fields
up to and including  $1/f^2$ terms and for
$F_X$ up to and including $1/f^3$ terms (due to its leading
contribution which is -$f$).
Doing so, in the final Lagrangian no kinetic mixing is present
at this order.
Using the expressions of the auxiliary fields, 
one then computes the $F$-part of the scalar potential 
of the Higgs sector, to find:
\medskip
\bea
V_F=\vert \mu\vert^2\,\Big[\vert h_1\vert^2+\vert
  h_2\vert^2\Big]
+\frac{\vert f+(B/f)\,h_1.h_2\vert^2}{
1+c_1\,\vert h_1\vert^2+c_2\,\vert  h_2\vert^2}+\cO(1/f^3)
\eea

\medskip\noindent
with $h_1.h_2\equiv h_1^0\,h_2^0-h_1^-\,h_2^+$ and $\vert
h_1\vert^2\equiv h_1^\dagger h_1=
h_1^{0\,*}h_1^0+ h_1^{-\,*} h_1^-$, etc.
One can work with this potential, however, for convenience,  if
$\vert c_{1,2}\vert \vert h_{1,2}\vert^2\ll~1$, we can approximate $V_F$
by expanding the denominator in a series of powers of these
coefficients. Our analysis below is then valid for
$\vert c_{1,2}\vert \vert h_{1,2}\vert^2\!\ll\!1$.
After adding the gauge contribution, we find the following result 
for the scalar potential of the Higgs sector:
\medskip
\bea
\label{potential0}
V&=&
f^2+
\big(\vert \mu\vert^2+m_1^2\big)\,\,
\vert h_1\vert^2+
\big(\vert \mu\vert^2 +m_2^2\big)
\vert h_2\vert^2
+\big(B\,h_1.h_2+h.c.\big)
\\[4pt]
&+&\!\!\!
\frac{1}{f^2}\,\Big\vert m_1^2\,\vert h_1\vert^2+m_2^2\,\vert h_2\vert^2+
B\,h_1.h_2\Big\vert^2
+\frac{g_1^2+g_2^2}{8}\,\Big[\vert h_1\vert^2-\vert h_2\vert^2\Big]^2
+\frac{g_2^2}{2}\,\vert h_1^\dagger\,h_2\vert^2
+\cO(1/f^3)\nonumber
\eea

\medskip\noindent
This is the full Higgs potential. The first term in the last
line  is a new term, absent in MSSM (generated by eliminating $F_X$ of
$X_{nl}$). Its  effects for phenomenology
will be analyzed later. The ignored higher order terms in $1/f$
involve non-renormalizable  $h_{1,2}^6$ interactions in $V$.

\section{New couplings in the Lagrangian.}\label{section3}

In this section we compute the new interactions induced by
Lagrangian (\ref{LL}), which are not present
in the MSSM. Many of the new couplings are actually dimension-four in fields,
with a (dimensionless)  $f$-dependent coupling.
The couplings are important in the case of a low SUSY
breaking scale in the  hidden-sector and a light gravitino scenario.
Some of the new couplings also involve  the goldstino field
 and  are relevant for phenomenology.

As mentioned earlier in Section~\ref{section2}, from the SUSY
 breaking part of the Lagrangian only terms up to and including
$1/f^2$  were kept in the total Lagrangian given by
equations (\ref{X}) to (\ref{gb}). After eliminating all terms proportional to
$F$-auxiliary fields of $X,H_i,Q,D^c,U^c,E^c,L$, one obtains
new couplings $\cL^{new}$ beyond those of the usual {\it
 onshell, supersymmetric} part of MSSM,
which are unchanged and not shown. One finds the onshell
Lagrangian
\bea
\cL^{new}\equiv \cL^{aux}_F+\cL^{aux}_D+\cL^{extra}_{m}+\cL^{extra}_{g}
\eea
Let us detail these terms. Firstly,
 \bea
 \cL^{aux}_F=\cL^{aux}_{F\, (1)}+\cL^{aux}_{F\, (2)}
 \eea
with
\medskip
\bea\label{auxLF1}
\cL^{aux}_{F\,(1)}
\!\!&=&\!\!
-\Big[
f^2+\big(m_1^2\vert h_1\vert^2+m_2^2\vert h_2\vert^2
+m^2_{\Phi}\,\vert\phi_\Phi\vert^2\big)\Big]
\nonumber\\
&-&
\Big[ B\,h_1.h_2
+A_u\,h_2.\phi_Q\,\phi_U
+A_d\,\phi_Q\,\phi_D .h_1
+A_e\,\phi_L\,\phi_E .h_1+\frac{1}{2}\,
m_{\lambda_i}\,\lambda_i\lambda_i+h.c.\Big]
\qquad
\eea

\medskip\noindent
recovering the usual MSSM soft terms
and the additional contributions:
\medskip
\bea\label{auxLF2}
\cL^{aux}_{F\,(2)}\!\!\!&=&\!\!\!\!
\Big\{\,
\frac{\opsi_X\opsi_X}{2\,f^2}
\Big[\mu
\big(m_1^2\!+\!m_2^2\big)\,h_1.h_2
\!-\!\big(m_1^2\!+\!m_Q^2\!+\!m_D^2\big) h_1.\phi_Q \phi_D
\!-\!\big(m_1^2\!+\!m_L^2\!+\!m_E^2\big) h_1.\phi_L \phi_E
\nonumber\\
&-&\!\!
\big(m_2^2+m_Q^2+m_U^2\big)\phi_Q\phi_U.h_2
\!+\!
\big(B\,h_2-A_d\,\phi_Q \phi_D-A_e\,\phi_L\,\phi_E\big)^\dagger
\big(\mu h_2-\phi_Q \phi_D-\phi_L\,\phi_E\big)
\nonumber\\[2.5pt]
&+&
\big(B\,h_1-A_u\,\phi_Q\,\phi_U\big)^\dagger
\big(\mu\,h_1-\phi_Q \,\phi_U\big)
+
\big(A_d\,\phi_D \,h_1-A_u\,h_2\,\phi_U\big)^\dagger
\big(\phi_D\,h_1-h_2\,\phi_U\big)
\nonumber\\
&+&A_d\,\big( \vert \phi_Q.h_1\vert^2+\vert \phi_E \,h_1\vert^2\big)
+A_u\,\vert h_2.\phi_Q\vert^2+A_e\,\vert \phi_L.h_1\vert^2\Big]
+h.c.\Big\}
-\frac{1}{f^2}\,
\Big\vert B\,h_1.h_2
\nonumber\\[-3pt]
&+&
\!\!\! A_u h_2.\phi_Q\,\phi_U
\!+\! A_d \phi_Q\,\phi_D .h_1
\!+\! A_e \phi_L\,\phi_E .h_1
\!+\! \frac{m_{\lambda_i}}{2}
\lambda_i\lambda_i
\!+\!
\big( m_1^2\vert h_1\vert^2
\!+\! m_2^2\vert h_2\vert^2
\!+\! m^2_{\Phi}\vert \phi_\Phi\vert^2\!\big)
\Big\vert^2
\nonumber\\[-1pt]
&-&
\frac{1}{f}\,\,\Big[
m_1^2\,\opsi_X\opsi_{h_1}\,h_1+
m_2^2\,\opsi_X\opsi_{h_2}\,h_2+
m^2_{\Phi}\,\opsi_X\opsi_\Phi\,\phi_\Phi+h.c.\Big]
+\cO(1/f^3)
\eea

\bigskip\noindent
A summation is understood over the SM group
indices: $i=1,2,3$ in the gaugino term
 and over $\Phi=Q,U^c,D^c,L,E^c$ in the mass terms;
appropriate contractions among $SU(2)_L$
doublets are understood for holomorphic products, when the order
displayed is relevant. The leading interactions $\cO(1/f)$
are those in the last line and are dimension-four in fields. Similar
couplings exist at  $\cO(1/f^2)$ and involve scalar and gaugino fields.
Yukawa matrices are restored in (\ref{auxLF2})
by replacing $\phi_Q\phi_D\ra \phi_Q\gamma_d\phi_D$,
$\phi_Q\phi_U\ra \phi_Q\gamma_u\phi_U$,
$\phi_L\phi_E\ra \phi_L\gamma_e\phi_E$, as already explained.

There are also new couplings
from terms involving the auxiliary components of the vector
superfields of the SM. Integrating them out
one finds:
\medskip
\bea
\cL_D^{aux}\!\!\!\!
&=&
\frac{-1}{2}\,
\Big[
\tilde D_1+ \frac{1}{4\, f^2} \,
\big( \,m_{\lambda_1}\,\psi_X\psi_X+h.c.\big)\,\tilde D_1
+\frac{1}{\sqrt 2 \,f}
\big(\,m_{\lambda_1}\,\psi_X\,\lambda_1+h.c.\big)\Big]^2
\nonumber\\
&+&\!\frac{-1}{2}\,\Big[
\tilde D_2^a +\frac{1}{4 \,f^2}\,
\big(m_{\lambda_2}\,\psi_X\psi_X+h.c.\big)\,\tilde D_2^a
+\frac{1}{\sqrt 2\,f}\big(
m_{\lambda_2}\,\psi_X\,\lambda_2^a+h.c.\big)\Big]^2
\nonumber\\
&+&\!\frac{-1}{2}\,\Big[
\tilde D_3^a +\frac{1}{4 \,f^2}\,
\big(m_{\lambda_3}\,\psi_X\psi_X+ h.c.\big)\,\tilde D_3^a\,
+\frac{1}{\sqrt 2\,f}\,\big(
m_{\lambda_3}\,\psi_X\,\lambda_3^a+h.c.\big)\Big]^2\!\!
+\cO(1/f^3)\qquad
\label{newD}
\eea

\medskip\noindent
with the notation:
\medskip
\bea\label{Dmssm}
\tilde D_{1}&=&
-\frac{1}{2}\,g_1\,
\big(- h_1^\dagger h_1+ h_2^\dagger h_2
+1/3\,\,\phi_Q^\dagger\phi_Q
-4/3\,\,\phi_U^\dagger\phi_U
+2/3\,\,\phi_D^\dagger\phi_D
-\phi_L^\dagger\phi_L
+2\,\phi_E^\dagger\phi_E\big)
\nonumber\\[-1pt]
\tilde D_{2}^a&=&
-\frac{1}{2}\, g_2\,
\big( h_1^\dagger\sigma^a h_1+h_2^\dagger\sigma^a
 h_2+\phi_Q^\dagger \sigma^a \phi_Q+\phi_L^\dagger\sigma^a
 \phi_L \big)
\nonumber\\[-1pt]
\tilde D_{3}^a&=&
-\frac{1}{2}\,g_3\,
\big(
\phi_Q^\dagger \,t^a\phi_Q
-
\phi_U^\dagger \,t^a\phi_U
-
\phi_D^\dagger \,t^a\phi_D\big)
\eea

\medskip\noindent
for the MSSM corresponding expressions; here $(t^a/2)$ are
the SU(3) generators.
From (\ref{newD}) one can easily read the
new, $f-$dependent couplings in the gauge sector,
absent in the MSSM.

The total Lagrangian also contains extra terms, not
proportional to  the auxiliary fields, and {\it not} present in the
MSSM. In the matter sector these are:
\medskip\noindent
\bea
\cL_m^{extra}\!\!\!\! &=&\!\!\frac{1}{4f^2}
\vert \partial_\mu(\psi_X\psi_X)\vert^2+
\Big(\frac{i}{2}\overline\psi_{X}\overline\sigma^\mu\,
\partial_\mu\psi_{X}+h.c.\Big)
\nonumber\\[-4pt]
&-&
\sum_{i=1}^2
\frac{m_i^2}{f^2}\,\Big\{\,
\overline\psi_X\overline\psi_{h_i}\,\psi_X\psi_{h_i}
\!+\!\Big[\,
\frac{i}{2} \,(\overline\psi_X\,
\overline\sigma^\mu\,\psi_X)\,(h_i^\dagger
\,\cD_\mu\,{h_i})
+\frac{i}{2}\vert h_i\vert^2\,
\overline\psi_X\,\overline\sigma^\mu\partial_\mu\psi_X+h.c.
\Big]\Big\}\nonumber\\[-3pt]
&-&
\Big[m_i^2\ra m_\Phi^2,
H_i\ra \Phi\Big]
%%%
+\bigg\{\,\,
\frac{B}{f}\,\,\Big[\,
\frac{1}{2\,f}\,\,\psi_X\psi_X \,\psi_{h_1}.\psi_{h_2}
-h_1.(\psi_{X}\psi_{h_2})-(\psi_X\psi_{h_1}).h_2\Big]
\nonumber\\[-3pt]
&+&
\frac{A_u}{f}\,\Big[\frac{1}{2\,f}
\psi_X\psi_X\,\big(
\,h_2.\psi_Q\,\psi_U
+\psi_{h_2}.\phi_Q\,\psi_U
+\psi_{h_2}.\psi_Q\,\phi_U\big)
-
\psi_X\,(h_2.\phi_Q\,\psi_U
+h_2.\psi_Q\,\phi_U
\nonumber\\[-2pt]
&+&
\psi_{h_2}.\phi_Q\,\phi_U)\Big]
+
 \Big[\frac{A_d}{f}\,\Big(\frac{1}{2\,f}\,\,
 \psi_X\psi_X\,(\psi_Q\,\psi_D.h_1
 +\,\phi_Q\,\psi_D.\psi_{h_1}+\psi_Q\,\phi_D.\psi_{h_1})\nonumber\\[-2pt]
 &-&\psi_X\,(\phi_Q\,\psi_D.h_1
 +\psi_Q\,\phi_D.h_1\!+\!\phi_Q\,\phi_D.\psi_{h_1})\Big)\!
 +\!(D\!\ra\! E, L\!\ra\! Q)\Big]
\!+\!h.c.\!\bigg\}\!+\!\cO(1/f^3).
\eea

\medskip\noindent
Note the presence of interactions that are dimension-four in
fields ($B/f \,h_1\psi_X \psi_{h_2}$, etc)
that can be relevant for phenomenology at low $f$.
There are also new couplings in the gauge sector
\medskip
\bea\label{lgextra}
\cL_g^{extra}&=&
\sum_{i=1}^3
\,\,\frac{m_{\lambda_i}}{2\,f}\,\,\Big[
\frac{\psi_X\psi_X}{-2 \,f}\,\Big(
2\,i\,\lambda^a\sigma^\mu\,\Delta_\mu\,\overline\lambda^a
-\frac{1}{2}\,F^a_{\mu\nu}\,F^{a\,\mu\nu} -\frac{i}{4}
\epsilon^{\mu\nu\rho\sigma}\,F^a_{\mu\nu}\,F^a_{\rho\sigma}
\Big)\qquad\qquad\qquad
\nonumber\\[2pt]
&-&\sqrt 2\,\psi_X\sigma^{\mu\nu}\lambda^a\,F_{\mu\nu}^a
\Big]_i+h.c.+\cO(1/f^3),
\eea

\medskip\noindent
with $i=1,2,3$ the gauge group index and
$\sigma^{\mu\nu}=i/4\,(\sigma^\mu\overline\sigma^\nu
-\sigma^\nu\overline\sigma^\mu)$.
 The new couplings of $\cL^{new}$ together
with the {\it onshell} part of the purely {\it supersymmetric}
part of the MSSM Lagrangian (onshell $\cL_0$ of (\ref{mssmsusy}))
gives the final onshell effective Lagrangian
of the model. From this, the full scalar potential is identified.

\section{Implications for the Higgs masses.}\label{section4}

Let us consider the Higgs scalar potential found in (\ref{potential0}) and
analyze the implications for the Higgs masses. From the neutral Higgs
part  of potential one finds the masses of the
CP even and CP odd Higgs fields. Exact values (in $1/f$) can be found
(see the Appendix), but since  
eq.\,(\ref{potential0}) is valid up to $1/f^4$ terms, it is sufficient to
 present the expressions of the Higgs masses that are
valid up to this order.
Firstly, at the minimum of the scalar potential one has:
\medskip
\bea\label{ttt}
 m_1^2-m_2^2&=&
\cot 2\beta\,\bigg[\,B+\frac{f^2}{v^2}
\frac{(-1+  \sqrt w_0)(-B+m_Z^2\,\sin 2\beta)}
{2\mu^2+m_Z^2 \cos^2 2\beta+ B\sin 2\beta}\,\bigg]
\nonumber\\[10pt]
m_1^2+m_2^2&=&
\frac{1}{\sin 2\beta}\,\bigg[
-B+\frac{f^2}{v^2}\frac{(-1+\sqrt w_0)(B+2\,\mu^2\,\sin 2\beta)}{
2\mu^2+m_Z^2 \cos^2 2\beta+ B\sin 2\beta}\,\bigg]
\eea
where
\bea
w_0\equiv
1-\frac{ v^2}{f^2}\,\big( 4\,\mu^2+2\,m_Z^2\,\cos^2
2\beta+2\,B\,\sin 2\beta\big)
\eea

\medskip\noindent
There is a second solution for $m^2_{1,2}$  at the minimum (with minus
in front of $\sqrt w_0$)
which however is not a perturbation of the MSSM one and not considered
below (since it brings a shift proportional to $f$ of the soft masses, 
which invalidates the expansion in $m_{1,2}^2/f$).
One finds  the following results
(upper sign for $m_h^2$ and lower sign for $m_H^2$):
\medskip
\bea
m_{h,H}^2&=&
\frac{1}{2}\,\,\Big[
m_Z^2+\frac{-2 \,B}{\sin
  2\beta}\mp \sqrt{w_1}\Big]
+
\frac{v^2}{32 f^2}\,\,
\Big\{
4\,B\,\Big[\,\,2 B+(4\mu^2+ 2
  m_Z^2\,\cos^2 2\beta)/\sin
  2\beta\Big]
\nonumber\\[-2pt]
&+&
4\,\,\,\Big[ \,\,2 \,B^2+ 8 \,\mu^4+ 2\,m_Z^2
(4 \mu^2+m_Z^2)\,\cos^2
  2\beta+8\,B\,\mu^2\sin
  2\beta\Big]\nonumber\\[-2pt]
&\mp&\!\!
\frac{\csc^2 2\beta}{\sqrt w_1}
\Big[-2 \,(B^2+4\mu^4)
  m_Z^2+4\mu^2 m_Z^4+m_Z^6
+8\, \big(2  \mu^4 m_Z^2- B^2\,(4\mu^2+m_Z^2)\big)\cos 4\beta
\nonumber\\[1pt]
&-& m_Z^2
  \,(6\,B^2+8\mu^4+4\mu^2
  m_Z^2+m_Z^4)\cos 8\beta
- 8\,B\,(B^2-8 \mu^4)\sin 2\beta
\nonumber\\[2pt]
&+& B(-8 B^2+16 \mu^2 m_Z^2 +m_Z^4)\sin
  6\beta+B m_Z^4\sin 10\beta\Big]\Big\}
+\cO(1/f^3)
\eea
with
\bea
w_1&=&\Big(m_Z^2+ \frac{-2\,B}{\sin
    2\beta}\Big)^2
-4\,m_Z^2\, \Big(\frac{-2\,B}{\sin
  2\beta}\Big)\,\cos^2 2\beta
\eea

\medskip\noindent
Further, the mass $m_A$ of the pseudoscalar Higgs
has a simple form (no expansion):
\medskip
\bea
m_A^2&=&\frac{-2\,B}{\sin 2\beta}\,\,\bigg\{\,
\frac{3}{4}+\frac{1}{4}\,\,\sqrt w_0
-\,\frac{v^2}{4\,f^2}\,B\,\sin 2\beta
\bigg\}
\eea

\medskip\noindent
and, as usual, the Goldstone mode has mass $m_G=0$.

\begin{figure}[t!h!]
\def\baselinestretch{1.}
\begin{center}
\begin{tabular}{cc|cr|}
\parbox{6.6cm}{
\subfloat[{\small $m_h$ in function of $\sqrt{f}$, $m_A$ parameter}]
{\includegraphics[width=6.6cm]{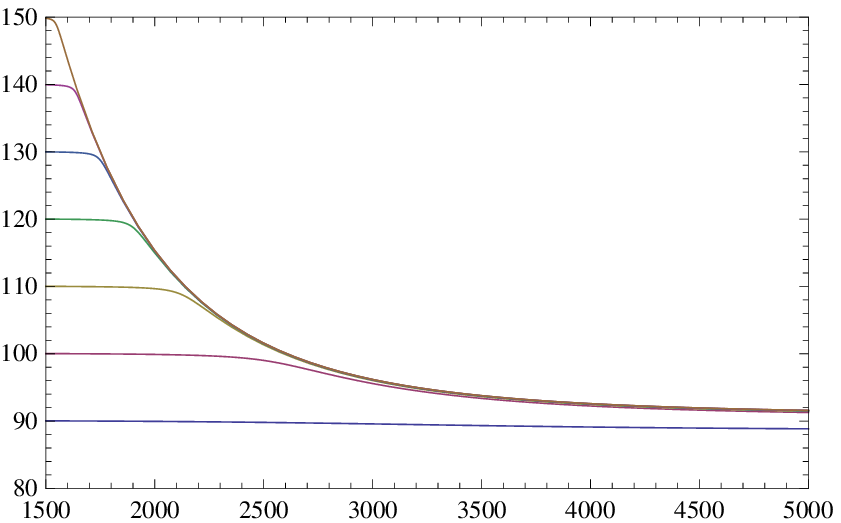}}}
\hspace{4mm}
\parbox{6.6cm}{
\subfloat[{\small $m_H$\,in function of $\sqrt{f}$, $m_A$ parameter}]
{\includegraphics[width=6.6cm]{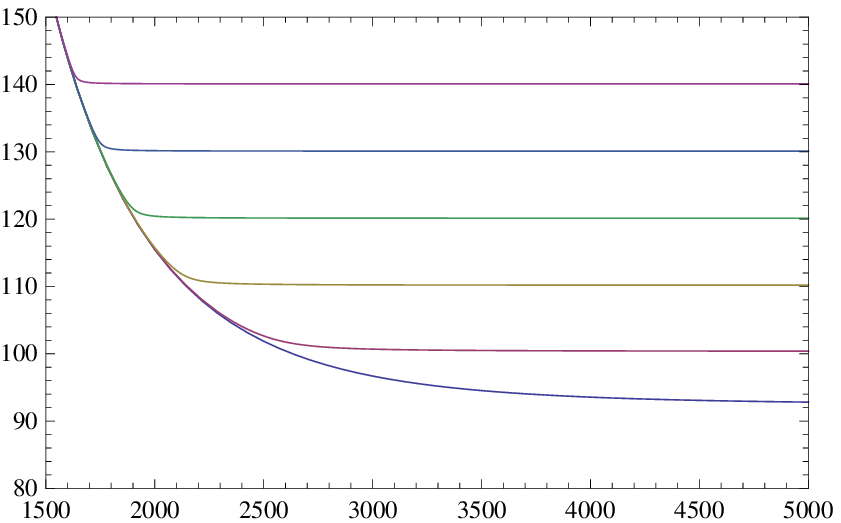}}}
\end{tabular}
\smallskip
\begin{tabular}{cc|cr|}
\parbox{6.6cm}{
\subfloat[{\small $m_h$ in function of $\sqrt{f}$, $\mu$ parameter}]
{\includegraphics[width=6.6cm]{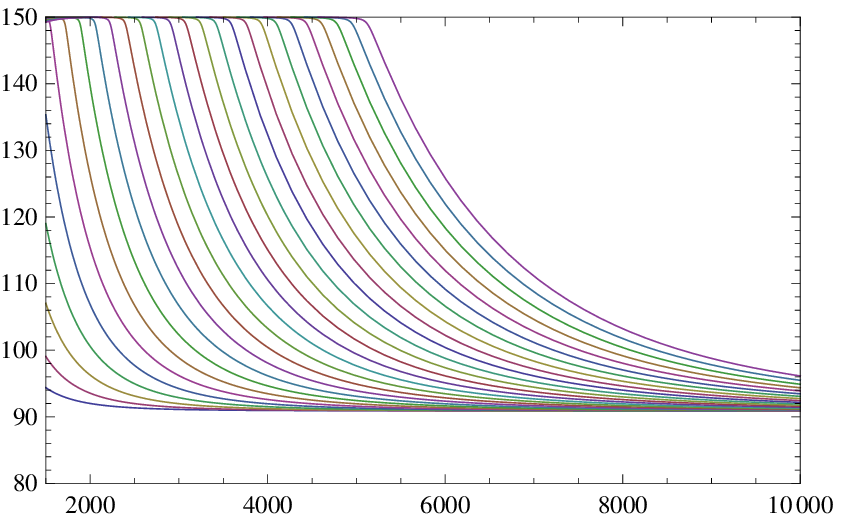}}}
\hspace{4mm}
\parbox{6.6cm}{
\subfloat[{\small $m_h$ in function of $\sqrt{f}$, $\mu$ parameter}]
{\includegraphics[width=6.6cm]{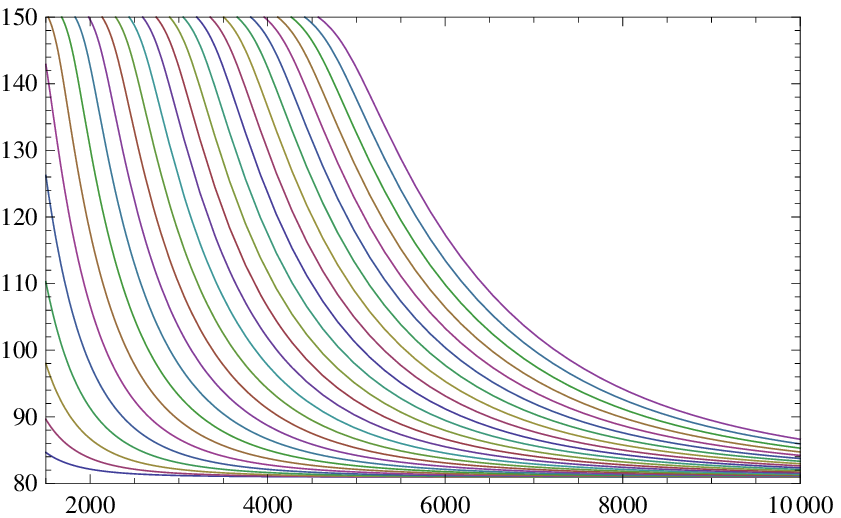}}}
\end{tabular}
\smallskip
\begin{tabular}{cc|cr|}
\parbox{6.6cm}{
\subfloat[{\small $c_1 v^2$ in function of $\sqrt{f}$}]
{\includegraphics[width=6.6cm]{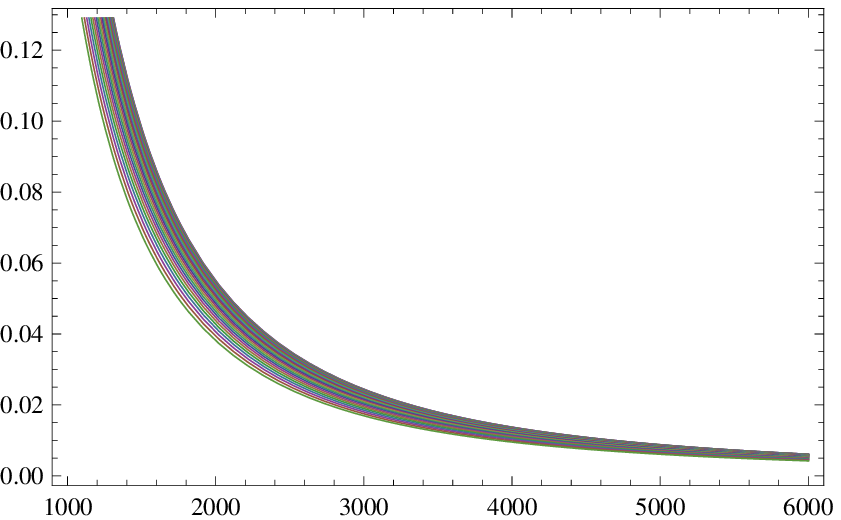}}}
\hspace{4mm}
\parbox{6.6cm}{
\subfloat[{\small $c_2 v^2$ in function of $\sqrt{f}$}]
{\includegraphics[width=6.5cm]{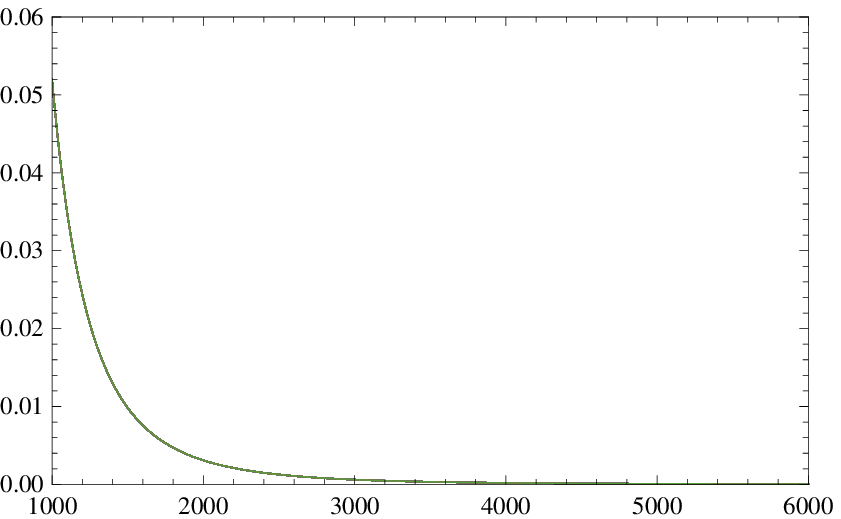}}}
\end{tabular}
\end{center}
\caption{{\protect\small
The tree-level Higgs masses (in GeV) and expansion
coefficients as functions of $\sqrt f$ (in GeV).
In (a), (b) $\mu\!=\!900$ GeV,
$\tan\beta\!=\!50$, $m_A$ increases upwards from $90$  to $150$ GeV
in steps of $10$ GeV.  
%Larger $m_A$ has little impact on $m_h$ for fixed $\sqrt f$. 
The increase of $m_h$ 
is significant even at larger $\sqrt f$, if one increases 
$\mu$, as seen in (c), (d).
In figs.~(c), (d), $m_A\!=\!150$ GeV and $m_h$ increases as
$\mu$ varies from 400 to 3000 GeV in steps of 100 GeV.
In (c) $\tan\beta\!=\!50$
while in (d) $\tan\beta\!=\!5$, showing a milder dependence on
$\tan\beta$ than in MSSM.
For $\tan\beta\!\geq\! 10$ there is little difference from (c).
In (e), (f) the expansion coefficients are shown, for
$m_A=[90,650]$ GeV with steps of $10$ GeV, $\mu\!=\!900$ GeV,
 $\tan\beta\!=\!50$; they are always less than
unity, even at larger values of $\sqrt f$ or $\mu$ shown in (c), (d), 
as required for a convergent expansion.}}
\label{higgs1}
\end{figure}

It is instructive to consider the limit of
large $u\equiv \tan\beta$, with $B<0$ fixed, when
\medskip
\bea
m_h^2\!\!\!&=&\!\!
\Big[m_Z^2+\cO (1/u)\Big]+\frac{v^2}{2\,f^2}\,\Big[
(2\,\mu^2+m_Z^2)^2+
\frac{4}{u}\,B\,(2\,\mu^2+m_Z^2)
+\cO(1/u^2)\Big]
+\cO(1/f^3)
\\[3pt]
m_H^2\!\!\!\!&=&\!\!\!
\Big[\frac{-2 B}{\sin 2\beta}\!+\!\cO(1/u)\Big]\!+\!
\frac{v^2\,B}{4\,f^2}\,\Big[
(2\,\mu^2+m_Z^2)\,u\!+\! 4\,B+
\frac{1}{u} (2\,\mu^2\!-\!11 m_Z^2)
\!+\!\cO(1/u^2)\Big]
\!+\!\cO(1/f^3)\nonumber
\eea

\medskip\noindent
which shows that a large $\mu$ can increase $m_h$ (decrease $m_H$).
However, for phenomenology it is customary to use $m_A$ as an input instead of
$B$, in which case the masses $m_{h,H}$ take the form
\medskip
\bea
m_{h,H}^2& =&
\frac{1}{2}\Big[ m_A^2+m_Z^2\mp \sqrt w\Big]
\pm \frac{v^2}{16 f^2}\frac{1}{\sqrt w}
\Big[
16 m_A^2\mu^4+4 \,m_A^2\,\mu^2\,m_Z^2+(m_A^2-8\,\mu^2)\,m_Z^4
\nonumber\\
&-&
2\,m_Z^6\pm
2\,(-2\,m_A^2\,\mu^2+8\mu^4+4\mu^2\, m_Z^2+m_Z^4)\,
\sqrt{w}+ m_A^2\,m_Z^4\cos 8\beta
\nonumber\\[2pt]
&+&
\!\!
m_A^4\,(m_A^2\!-\! 8\mu^2\!-\! 3 m_Z^2)\sin^2 2\beta
\!+\!
\cos 4\beta\,\big[-2 m_Z^2 \,(8\mu^4\! +\! 4\mu^2\,m_Z^2\! +\!
m_Z^4\!-\! m_A^2 (6\mu^2\!+\! m_Z^2))
\nonumber\\[2pt]
&\pm &
2 \,(2\,m_A^2\mu^2+4\mu^2 m_Z^2 +m_Z^4)\,\sqrt w
-m_A^2 (m_A^2+5 \,m_Z^2)\,\sin^2 2\beta\,
\big]\,\Big]+\cO(1/f^3)
\eea

\medskip\noindent
where the first term (bracket) is just the MSSM contribution.
The upper (lower) signs correspond to $m_h$ ($m_H$) and
$w=(m_A^2+m_Z^2)^2-4\,m_A^2\,m_Z^2\,\cos^2 2\beta$.
At large $\tan\beta$ with $m_A$ fixed one finds\footnote{
In (\ref{lim}) $m_A>m_Z$ is assumed, otherwise just exchange $m^2_{h}$
with $m_H^2$.}  (with $u\equiv \tan\beta$)
\medskip
\bea\label{lim}
m_h^2&=&\Big[m_Z^2+\cO(1/u^2)\Big]
+\frac{v^2}{2\,f^2}\,\Big[(2 \,\mu^2+m_Z^2)^2+\cO(1/u^2)\Big]
+\cO(1/f^3)
\nonumber\\
m_H^2&=&\Big[m_A^2+\cO(1/u^2)\Big]
+\frac{1}{f^2}\,\cO(1/u^2)
+\cO(1/f^3)
\eea
In this limit the increase of $m_h$  is driven by a 
large $\mu$ and apparently is of SUSY origin,
but the quartic Higgs couplings giving this effect
involved combinations of soft masses (see (\ref{potential0})).
 These soft masses combined to give, at the EW minimum, 
the $\mu$-dependent increase in
(\ref{lim})\footnote{See also
 $\lambda$ of (\ref{lam}) evaluated at EW minimum, 
$\delta=0$, $\tan\beta\!\ra\! \infty$:
$\lambda\ra (1/2 v^2)[m_Z^2+v^2(m_Z^2+2\mu^2)^2/(2f^2)]$.}.

Some simple numerical examples are relevant for the size of the corrections
to the Higgs masses, relative to their MSSM values. The largest
correction to $m_h$
for large $\tan\beta$ is  dominated by $\mu$ and $f$ (eq.\,(\ref{lim})).
For example, if $(\mu/\sqrt f)^2=(1/2.25)^2\approx 1/5$, $v=246$ GeV,
with $\mu=900$ GeV then $\sqrt f=2$ TeV, giving $m_h=114.4$ GeV.
Other  examples are: ($\mu=1.2$ TeV, $\sqrt f=2.7$ TeV) and
($\mu=2.6$ TeV, $\sqrt f=6$ TeV), leading to $m_h=114.4$ GeV
(with $(\mu/\sqrt f)^2\approx 1/5$).
Smaller  $\mu\approx 600$ GeV can still allow $m_h$ just above the LEP bound
if $\sqrt f=1.35$ TeV, for similar value for $(\mu/\sqrt f)^2=1/5$
and for the rest of the parameters.
This  shows that one can have a {\it classical} value of $m_h$
near or marginally above the LEP bound and larger than the classical
MSSM value ($=m_Z$).
The plots in Figure~\ref{higgs1} illustrate
better the value of  $m_h$ and $m_H$ for various values of $\sqrt f$.
For $\sqrt f$ in the region of $1.5$ TeV to 7 TeV the LEP bound is
satisfied for $m_h$, while at larger $\sqrt f$ the MSSM case is recovered.
By varying  $\sqrt f$ our results can interpolate
between low and high scale (in the hidden sector) SUSY breaking.
Quantum corrections  increase $m_h$ further, just as in the MSSM.

Regarding the usual MSSM tree-level flat direction $\vert
h_1^0\vert=\vert h_2^0\vert$ one can show that the potential in this
direction can have a minimum for the case (not considered in MSSM)
of $m_1^2+m_2^2+2\vert \mu\vert^2< 2\vert B\vert$, equal to
$V_m=f^2-(1/4)f^2 (m_1^2+m_2^2+2\vert\mu\vert^2+2 B)^2/(m_1^2+m_2^2+B)^2.$
Compared to the usual MSSM minimum, the former can be situated above it
only for values of $f$ which do not comply with the original
assumptions of $m_{1,2}^2, \vert B\vert<f$.  On the other hand,
the case with $V_m$ situated below the MSSM minimum does not allow
one to recover the MSSM ground state in the decoupling limit of large
$f$, and in conclusion the ``flat'' direction is not of  physical
interest here.

\section{Other phenomenological implications.}\label{section5}

\subsection{Fine-tuning of the electroweak scale}

The increase of $m_h$, at the classical level, beyond the MSSM
tree-level bound ($m_Z$) and the
presence of the new quartic couplings of the Higgs fields
also have   implications for the fine tuning.
In the MSSM the smallness of the effective
quartic coupling $\lambda$ (fixed by the gauge sector) is at the
origin of an increased amount of fine tuning of the electroweak scale
for large soft masses.
For soft masses significantly larger than the electroweak (EW) scale,
(also needed  to
increase the MSSM value for $m_h$ above LEP bound via quantum corrections),
fine tuning increases rapidly\footnote{Two-loop 
MSSM fine tuning \cite{Cassel:2010px} 
is minimized at $m_h\sim 115$ GeV (consistent with EW and
dark matter constraints); however, beyond this value, fine tuning
increases {\it exponentially} with $m_h$.} and may become a potential
problem (sometimes referred to  as the ``little hierarchy'' problem).
Let us see why in the present model
this problem is alleviated. One can write $v^2=-m^2/\lambda$ where
\bea\label{lam}
\lambda &\equiv& \frac{g_1^2+g_2^2}{8}
\Big[\cos^2 2\beta+\delta \sin^4 \beta\Big]
+\frac{1}{f^2}\,\Big\vert
m_1^2\cos^2\beta+m_2^2 \sin^2 \beta +(1/2) \,B\,\sin  2\beta
\Big\vert^2\nonumber\\
m^2&\equiv&  (\vert \mu\vert^2+m_1^2)\cos^2\beta+(\vert \mu\vert^2 +
m_2^2) \sin^2 \beta + \,B\,\sin 2\beta\
\eea
The first term in $\lambda$ is due to MSSM only, while the
second one, which is positive, is due to the new quartic Higgs terms
in (\ref{potential0}).
Here $\delta$ accounts for
the top/stop quantum effects to $\vert h_2\vert^4$ term in the
potential, which becomes $(1+\delta)\,(g_1^2+g_2^2)/8\,\vert
 h_2\vert^4$; usually
 $\delta\sim \cO(1)$ (ignoring couplings other than top
 Yukawa). This quantum effect is only included for a comparison to
the new quartic Higgs term. The important point to note
is that a larger  $\lambda$ gives a suppression in the
fine tuning measure $\Delta$:
%\medskip
\bea
\Delta=\frac{\partial \ln v^2}{\partial \ln p}
=\frac{\partial \ln (-m^2/\lambda)}{\partial \ln p},
\qquad p=A, B, m_0^2, \mu^2, m_{\lambda_i}^2.
\eea
%\medskip\noindent
Here $p$ is an MSSM parameter 
with respect to which fine tuning is evaluated.
In  the large $\tan\beta$ limit, the fine tuning of the electroweak
scale becomes (see the Appendix in \cite{Cassel:2009ps}):
%\medskip
\bea
\Delta=-\frac{
(\vert \mu\vert^2+m_2^2)'}{v^2\,{m_2^4}/{f^2}+(1+\delta)\,
{m_Z^2}/{2}}+\cO(1/\tan\beta),\qquad
(\vert \mu\vert^2+m_2)'\equiv \frac{\partial (\vert
  \mu\vert^2+m_2^2)}{\partial \ln p}
 \eea
%\medskip\noindent
For small $\tan\beta$ a similar result is obtained in which one
replaces $m_2$ by $m_1$.
The first term in denominator comes from the new correction to
the effective quartic coupling $\lambda$.
Larger soft masses $m_{1,2}$ increase $\lambda$  and
this can  actually reduce fine tuning, see the denominator in $\Delta$.
Therefore, in this case heavier superpartners do not
necessarily bring an increased fine tuning amount (as it usually
happens in the MSSM). The only limitation here is the size of the ratio
$m^2_{1,2}/f\leq 1$ for convergence of the nonlinear formalism. In
the limit this coefficient approaches its upper limit (say $\sim 1/3$),
the two contributions in the denominator have comparable size (for
$\delta\sim 1$ and $v=246$ GeV) and fine tuning is reduced
by a factor $\approx 2$ from that in the absence of the new term in
the denominator ({\it i.e.} the MSSM case).

\subsection{Limiting cases and loop corrections.}

Some interesting limits of our ``non-linear''  MSSM model are worth
considering. Firstly, in the limit of large $f$ 
({\em i.e.} large SUSY breaking scale in the hidden sector) 
and with $m_{1,2}, B$ {\it fixed}, the new quartic term in (\ref{potential0})
vanishes, while the usual explicit soft SUSY breaking terms specific
to the Higgs sector remain. This is just the MSSM case. All other
couplings suppressed by inverse powers of $f$ are negligible in this
limit.
Another limiting case is that of very small $f$. For our analysis to
be valid, one needs to satisfy the condition
$B,\, m_{1,2}^2\leq f.\,\,$
When $f$ reaches this minimal bound, the new quartic couplings in
(\ref{potential0}), not present in the MSSM, increase and eventually
become  closer to unity.
The analysis is then less reliable and additional effective
contributions in the Lagrangian,
suppressed by higher powers like $1/f^4$ and beyond, may  become
relevant for SUSY breaking effects.

Finally, one remark regarding the calculation of
radiative corrections using (\ref{potential0})
and the electroweak symmetry breaking (EWSB).
In our case EWSB was assumed to take place 
by appropriate values of $m_{1,2}^2, B$. However, 
 the same EWSB mechanism as in the
MSSM is at work here, via quantum corrections to 
these masses, which near the EW
scale turn $m_2^2+\mu^2$ negative and trigger radiative EWSB.
Indeed, if the loops of the MSSM states  are cut off as usual 
at the high GUT scale (well above $\sqrt f$) and with the new Higgs 
quartic  couplings regarded as an {\it effective}, classical operator,  
radiative EWSB can take place  as in the MSSM. A similar example is
the case of a MSSM Higgs sector extended  with additional
 effective operators of dimension $d=5$ such as 
$(1/M)\int d^2\theta (H_1 H_2)^2$ giving a dimension-four 
(in fields) contribution to the scalar potential 
$V\supset (\mu/M) h_1 h_2\,(\vert h_1\vert^2+\vert h_2\vert^2)$;
this is regarded as an effective operator and radiative EWSB is
 implemented 
 as  in the MSSM, see for example \cite{Dine:2007xi,Cassel:2009ps}.
The advantage in our case is that no ``new physics'' (scale $M$) is
introduced in the visible sector. In both cases, the new scale $M$ and
our scale $\sqrt f$ have comparable values, because in both cases the
increase of $m_h$ above the LEP bound is done via couplings
depending on the ratio $(\mu/M)$ 
\cite{Antoniadis:2009rn,Dine:2007xi,Cassel:2009ps,Blum:2009na,Carena:2009gx}
or $(\mu/\sqrt f)$, respectively.

It is interesting to remark that that the
loop corrections induced by the (effective) quartic couplings 
proportional to $1/f^2$ in eq.(\ref{potential0}), can be under control at
large $f$. Indeed, the loop integrals this coupling induces can be 
quadratically divergent and are then cut-off at momentum 
$p^2\leq f$; but the loop effects 
come with  a coupling factor that behaves like $1/f^2$, so overall  they
will be suppressed like $1/f$ and can then be under
control even at large f (for a discussion of loop corrections involving the
goldstino, see \cite{JBAF}).

\subsection{Invisible decays of Higgs and $Z$ bosons.}\label{invisible}

Let us analyze some implications of the interactions involving the
goldstino field, described by the Lagrangian found above.
An interesting possibility, for a light enough neutralino, is the decay
of the neutral higgses into a goldstino and the lightest
neutralino $\chi_1^0$ (this is the NLSP, while goldstino is the LSP).
The coupling Higgs-goldstino-neutralino is only suppressed by $1/f$. 
It arises from the following terms in $\cL^{new}$ and from the terms
in the {\it onshell, SUSY} part
of usual MSSM Lagrangian (\ref{mssmsusy}), hereafter denoted
 $\cL_0^{onshell}$:
\medskip
\bea\!\!\!
\cL^{new}+\cL^{onshell}_0\!\!\!
&\supset&\!\!\!\!
-\frac{1}{f}\,\,
\Big[
m_1^2 \,\,\psi_X\psi_{h_1^0}\,h_1^{0\,*}
+m_2^2 \,\,\psi_X\psi_{h_2^0}\,h_2^{0\,*}
\Big]
-\frac{B}{f}\,\,\Big[\psi_X\psi_{h_2^0}\,h_1^0
+\psi_X\psi_{h_1^0}\,h_2^0\Big]
\nonumber\\
&-&
\frac{1}{f}\sum_{i=1,2}\,\frac{m_{\lambda_i}}{\sqrt 2}\,
\tilde D_i^a\,\psi_X\lambda_i^a
-
\frac{1}{\sqrt 2}
\Big[g_2\lambda_2^3-g_1\lambda_1\Big]
\Big[h_1^{0\,*}\psi_{h_1^0}
-
h_2^{0\,*}\psi_{h_2^0}\Big]
+h.c.\,\,\,\,\,\,\,\,
\label{tt}
\eea

\medskip\noindent
The last term (present in the MSSM) also brings
a goldstino interaction. This is possible through the goldstino
components of the higgsinos $\psi_{h_{1,2}^0}$
 and EW gauginos $\lambda_{1,2}$.
The goldstino components  are found via the
equations of motion, after EWSB, to give (see also \cite{SK}):
\medskip
\bea\label{Gcom}
\mu\,\psi_{h_1^0}&=&\frac{1}{f\,\sqrt 2}\,\Big(
-m_2^2\,\,v_2-B\,v_1 -\frac{1}{2}\,\,v_2\,\,
\langle g_2 D_2^3-g_1 D_1\rangle\Big)\,\psi_X+\cdots
\nonumber\\[-2pt]
\mu\,\psi_{h_2^0}&=&\frac{1}{f\,\sqrt 2}\,\Big(
-m_1^2\,\,v_1-B\,v_2 +\frac{1}{2}\,\,v_1\,\,
\langle g_2 D_2^3-g_1  D_1\rangle\Big)\,\psi_X+\cdots
\nonumber\\
\lambda_1&=&
\frac{-1}{f\,\sqrt 2}\,\langle D_1\rangle\,\,\psi_X+\cdots,\qquad
\lambda_2^3=\frac{-1}{f\,\sqrt 2}\,\langle D_2^3\rangle\,\,\psi_X+\cdots
\eea

\medskip\noindent
which can be further simplified by using the MSSM minimum
conditions in the terms multiplied by $1/f$ (allowed in this
approximation).
As a consistency check we also showed that the determinant of the
neutralino mass matrix  (now a $5\times 5$ matrix, to include the
Goldstino)  vanishes up to corrections of order $\cO(f^{-4})$.
This is consistent with our approximation for the Lagrangian,
and verifies the existence of a massless  Goldstino
(ultimately ``eaten'' by the gravitino).
Using (\ref{tt}) and (\ref{Gcom}), one finds after
some calculations (for previous calculations of this decay
see \cite{Djouadi:1997gw,hgh,Dimopoulos:1996yq}):

\bea
\cL^{new}+\cL^{onshell}_0
\supset
-\frac{1}{f\sqrt 2}\,\sum_{j,k=1}^4\Big[\,
\psi_X\,\chi_j^0\,H^0 \,\delta_k\,\cZ_{jk}^*
+
\psi_X\,\chi_j^0\,h^0 \,\delta_k^\prime\,\cZ^*_{jk}\Big]
+h.c.
\eea
where
\bea
\delta_1&=&\,\,\,\,\,m_Z\,\sin\theta_w\,
\big[m_{\lambda_1}\cos(\alpha+\beta)+\mu\sin(\alpha-\beta)\big],
\nonumber\\
\delta_2&=&-
m_Z\cos\theta_w\,
\big[m_{\lambda_2}\cos(\alpha+\beta)+\mu\sin(\alpha-\beta)\big],
\nonumber\\
\delta_3&=&
-m_A^2\sin\beta\,\sin(\alpha-\beta)-\mu^2\cos\alpha
\nonumber\\
\delta_4&=&\,\,\,\,
m_A^2\cos\beta\,\sin(\alpha-\beta)-\mu^2\sin\alpha,
\qquad
\delta_i^\prime=\delta_i\Big\vert_{\alpha\ra \alpha+\pi/2}
\eea

\medskip\noindent
 $\cZ$ is the matrix that diagonalizes the MSSM
neutralino mass matrix\footnote{The exact form of $M$ is:
$M_{11}=m_{\lambda_1}$, $M_{12}=0$,
$M_{13}=-m_Z\cos\beta\sin\theta_w$,
$M_{14}=m_Z\sin\beta\sin\theta_w$,
$M_{21}=0$,
$M_{22}=m_{\lambda_2}$,
$M_{23}=m_Z\cos\beta\cos\theta_w$,
$M_{24}=-m_Z\sin\beta\cos\theta_w$,
$M_{33}=0$,
$M_{34}=\mu$, $M_{44}=0$, also $M_{ij}=M_{ji}$. Note
the sign of $\mu$ related to our
definition of the holomorphic product of SU(2) doublets.
With this notation, in the text $\chi_j^0=\cZ_{jk}\,\xi_k$,
with $\xi_k^T\equiv (\lambda_1,\lambda^3_2,\psi_{h_1^0},\psi_{h_2^0})$.}:
 $M_d^2=\cZ\,M\,M^\dagger\,\cZ^\dagger$, and can be easily evaluated
 numerically (see \cite{ElKheishen:1992yv} for its
 analytical expression).
Further $H^0, h^0$ are Higgs mass eigenstates (of mass $m_{h,H}$
 computed earlier) and
$h_i^0=1/\sqrt{2}\,\,(v_i+h_i^{0\, \prime}+i \sigma_i)$ with $\langle
h_i^{0\,\prime}\rangle=0$,  $\langle \sigma_i\rangle=0$; the
 relation of $H^0, h^0$ to $h_{1,2}^{0\,'}$ is a rotation, which in
 this case can be just that of the MSSM (due to extra $1/f$
 suppression in the coupling\footnote{The relation is $h_1^{0\,'}=H^0
\cos\alpha-h^0 \sin \alpha$, and  $h_2^{0\,'}=H^0
\sin\alpha+h^0 \cos \alpha$.}).
The angle $\alpha$ is
\bea
\tan 2\alpha=\tan 2\beta\,\,
\frac{m_A^2+m_Z^2}{m_A^2-m_Z^2},\qquad
-\pi/2 \leq \alpha\leq 0
\eea

\medskip\noindent
If the lightest neutralino is light enough, $m_{\chi_1^0}< m_h$,
 then $h^0, H^0$ can decay into it and a goldstino
 which has a  mass of order $f/M_{Planck}\sim 10^{-3}$ eV; if this is
 not the case, the decay of neutralino into $h^0$ and goldstino
takes place,  examined in \cite{Dimopoulos:1996yq}.
In the former case, the partial decay rate is
\medskip
\bea
\Gamma_{h^0\ra \chi_1^0\,\psi_X}=
\frac{m_h}{16\,\pi\,f^2}\,\,
\Big\vert \sum_{k=1}^4
\delta'_k\,\cZ_{1k}\Big\vert^2\,
\,\bigg(1-\frac{m_{\chi_1^0}^2}{m_{h^0}^2}
\,\bigg)^2
\eea

\medskip\noindent
The partial decay rate has corrections coming from both
higgsino ($\cZ_{13}$, $\cZ_{14}$) and gaugino fields  ($\cZ_{11}$,
$\cZ_{12}$),  since they both acquire a goldstino
component, see eqs.~(\ref{Gcom}).  The gaugino correction arises after 
gaugino-goldstino mixing, SUSY and EW symmetry breaking,
(as shown by $m_{\lambda_i}$, $m_Z$ dependence in  $\delta_k'$)
and was not included in previous similar studies
\cite{Djouadi:1997gw,hgh,Dimopoulos:1996yq}.

\begin{figure}[t!]
\def\baselinestretch{1.}
\begin{center}
\subfloat[$\Gamma_{h^0\ra\chi\psi_x}$
 in function of $\sqrt{f}$.]
{\includegraphics[width=6.7cm]{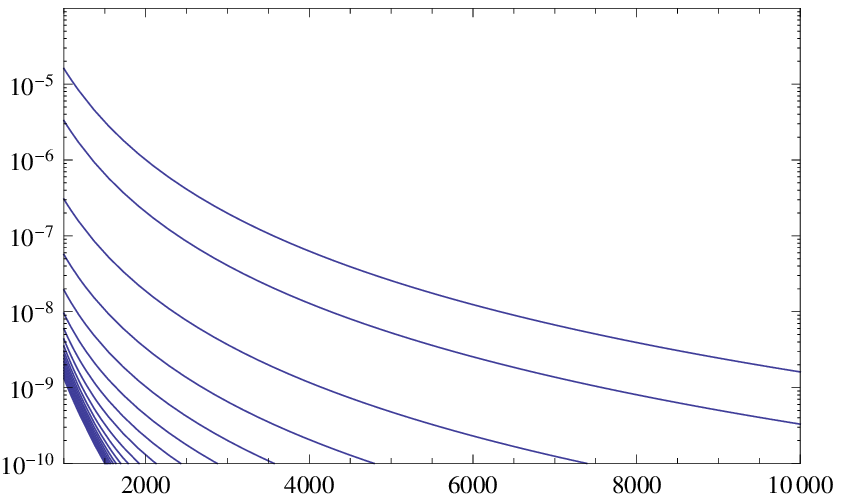}}
\hspace{7mm}
 \subfloat[$\Gamma_{h^0\ra\chi\psi_x}$
 in function of $\sqrt{f}$.]
 {\includegraphics[width=6.7cm]{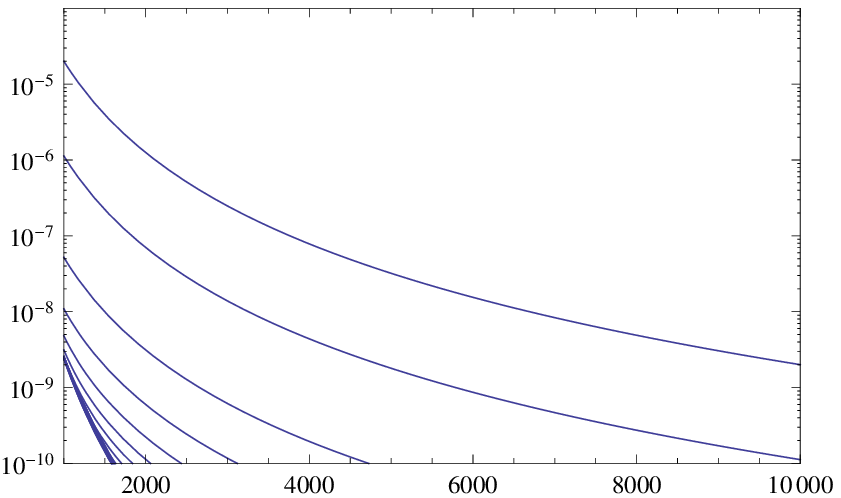}}
\end{center}
\caption{{\protect\small
The partial decay rate of $h^0\ra \psi_X\chi_1^0$ for
(a): $\tan\beta=50$, $m_{\lambda_1}=70$ GeV, $m_{\lambda_2}=150$ GeV,
$\mu$ increases from 50 GeV (top curve) by a step 50 GeV,
$m_A=150$ GeV. Compare against Figure~\ref{higgs1} (c) corresponding
to a similar range for the parameters.
At larger $\mu$, $m_h$ increases, but the partial decay rate decreases.
Similar picture is obtained at low $\tan\beta\sim 5$.
 (b):  As for (a) but with  $\tan\beta=5$. Compare against
 Figure~\ref{higgs1} (d).
 Note that the total SM decay rate, for
$m_h\sim 114$ GeV, is of order $10^{-3}$, thus the branching ratio in
the above cases becomes comparable to that of SM Higgs going into
$\gamma\gamma$  (see Figure 2 in \cite{Djouadi:1997yw}).}}
\label{decayplots}
\end{figure}

The partial decay rate is presented in Figure~\ref{decayplots}
for various values of $\mu$, $m_A$ and $m_{\lambda_{1,2}}$
which are parameters of the model. A larger decay rate requires
a light $\mu\sim \cO(100)$ GeV, when the neutralino $\chi_1^0$ has
a larger higgsino component. At the same time an increase of $m_h$
above the LEP bound requires a larger value for $\mu$,
close to $\mu\approx 700$ GeV if $\sqrt f\approx 1.5 $ TeV, and
$\mu\approx 850$ GeV if $\sqrt f\approx 2$ TeV,
see Figure~\ref{higgs1} (c).  The results in Figure~\ref{decayplots}
show that the partial decay rate can be significant ($\sim 3\times
 10^{-6}$ GeV), if we recall that the total SM Higgs decay rate 
(for $m_h\approx 114$ GeV) is about $3\times 10^{-3}$ GeV,
with a  branching ratio of $h^0\ra \gamma\gamma$ of
 $2\times 10^{-3}$, (Figure 2 in \cite{Djouadi:1997yw}). 
Thus the branching ratio of the process can be
close to that of SM  $h^0\ra \gamma\gamma$.
The decay is not very sensitive to $\tan\beta$ (Figure~\ref{decayplots} (b)),
due to the extra contribution (beyond MSSM) from the quartic Higgs coupling.
It would be interesting to analyze the above decay rate
at the one-loop level, for a more careful comparison to SM Higgs
decays rates.

An interesting coupling that is also present in the $1/f$ order is
that of goldstino to $Z_\mu$ boson and to a neutralino. Depending on
the relative mass relations, it can bring about a decay of $Z_\mu$
($\chi_j^0$) into $\chi_j^0$ ($Z_\mu$) and a goldstino, respectively. The relevant
terms are
\medskip
\bea\label{sqw}
\cL^{new}+\cL_0^{onshell}&\supset&
-
\frac{1}{4}\,\,\overline \psi_{h_1^0}\overline\sigma^\mu\psi_{h_1^0}
\,\,(g_2 V_2^3-g_1\,V_1)_\mu\,
+
\frac{1}{4}\,\,\overline \psi_{h_2^0}\overline\sigma^\mu\psi_{h_2^0}
\,\,(g_2 V_2^3-g_1\,V_1)_\mu\,\Big\}
\nonumber\\
&-&
\sum_{i=1}^{2}
\frac{m_{\lambda_i}}{\sqrt 2\,\,f}\,\psi_X\,\sigma^{\mu\nu}\,
\lambda_i^a\,F_{\mu\nu,\,i}^a+h.c.
\eea

\medskip\noindent
where the last term was generated in (\ref{lgextra}) ($i$ labels the
gauge group).
Since the higgsinos
acquired a goldstino component ($\propto \psi_X/f$)
via mass mixing, the first line above induces additional $\cO(1/f)$
couplings of the higgsino to goldstino and to
$Z_\mu=(1/g)\,(g_2 V_2^3-g_1 \,V_1)_\mu$ with $g^2=g_1^2+g_2^2$.
After some calculations one finds the coupling $Z_\mu\,\chi_j^0\,\psi_X$:
\medskip\noindent
\bea\label{rrr}
\cL^{new}\!\!+\!\cL_0^{onshell}\!\!
=\!
\frac{1}{f\sqrt 2}\!
\sum_{j=1}^4\bigg[
\overline\psi_X\overline\sigma^\mu\,\chi_j^0\,Z_\mu\,
\big(\mu\,m_Z\,w_j\! -\! m_Z^2 v_j\big)\!-\!
\opsi_X (\overline\sigma^\mu\partial^\nu\!-\!
\overline\sigma^\nu\partial^\mu)
\chi_j^0 Z_{\mu\nu} v_j\bigg]\!+\!hc
\eea

\medskip\noindent
where
\bea
w_j=\cos\beta\,\cZ_{j4}^*-\sin\beta\,\cZ_{j3}^*,\quad
v_j=-\sin\theta_w\,\cZ_{j1}^*+\cos\theta_w\,\cZ_{j2}^*,\quad
Z_{\mu\nu}=\partial_\mu Z_\nu-\partial_\nu Z_\mu
\eea

\medskip\noindent
If $m_{\chi_1^0}$ is lighter than $Z_\mu$ then a decay of the latter
into $\chi_1^0+\psi_X$ is possible.
The decay rate of this process is (with $j=1$):
\medskip
\bea
\Gamma_{Z\ra \psi_X\chi_j^0}&=&
\frac{m_Z^5}{32\pi f^2}
\Big[
\zeta_1\vert w_j\vert^2 +\zeta_2\,\vert v_j\vert^2
+\zeta_3\,(w_j\,v_j^*+w_j^*\,v_j)\Big]
\Big(1-\frac{m_{\chi_j}^2}{m_Z^2}\Big)^2
\label{r4}
\eea

\medskip\noindent
with $\zeta_1=2 (2+r^2)\,\mu^2/m_{Z}^2$,
$\zeta_2= 2(8+r^2)(1+2 r^2)$,
$\zeta_3=-2(4+5 r^2) \mu/m_Z$ where $r=m_{\chi_j}/m_Z$
(in (\ref{rrr}) and subsequent one can actually replace  $\mu$ by $m_{\chi_j}$ 
and $w_j\ra w_j^*$, with $\cZ_{j4}\leftrightarrow \cZ_{j3}$).

The decay rate  should be within
the LEP error for $\Gamma_Z$, which is
$2.3$ MeV \cite{pdg} (ignoring theoretical uncertainties
 which are small).
 From this, one finds a lower bound for $\sqrt f$, which can be as
 high as $\sqrt f\approx 700$ GeV for the parameter space considered
 previously in Figure~\ref{higgs1}, while  generic values are  $\sqrt
 f\sim \cO(400)$ GeV.
Therefore the results for the  increase of $m_h$, that needed
a value for $\sqrt f$ in the TeV region, escape this constraint.
This constraint does not apply if the lightest neutralino has a mass
larger than $m_Z$, when  the opposite decay ($\chi_j\ra Z\,\psi_X$)
takes  place (this can be arranged for example by a larger $m_{\lambda_1}$).

There  also exists the interesting possibility of an invisible decay 
of $Z_\mu$ gauge boson into a pair of goldstino fields, that we review
here \cite{Luty:1998np,SK,Brignole:2003cm}. This is induced by the
following terms in the Lagrangian, after the Higgs field acquires a
VEV:
\medskip
\bea\label{coupl}\!\!\!\!\!\!
\cL^{new}+\cL^{onshell}_0\!\!\!\!
&\supset &\Big\{
\frac{1}{4\,f^2}\,\,
\overline\psi_X\overline\sigma^\mu\psi_X\,\,
(g_2 V_2^3-g_1\,V_1)_\mu\,(m_1^2 \,v_1^2/2-m_2^2 \,v_2^2/2)
\nonumber\\
&-&
\frac{1}{4}\,\,\overline \psi_{h_1^0}\overline\sigma^\mu\psi_{h_1^0}
\,\,(g_2 V_2^3-g_1\,V_1)_\mu\,
+
\frac{1}{4}\,\,\overline \psi_{h_2^0}\overline\sigma^\mu\psi_{h_2^0}
\,\,(g_2 V_2^3-g_1\,V_1)_\mu\,\Big\}+h.c.
\eea

\bigskip\noindent
With (\ref{Gcom}) and (\ref{coupl})
one finds the coupling of $Z$ boson to
a pair of goldstinos:
\medskip\noindent
\bea
\cL^{new}+\cL^{onshell}_0
\supset
\frac{m_Z^2}{4\,f^2}\,\,
\overline\psi_X\,\overline\sigma^\mu\,\psi_X\,\,Z_\mu\,
\langle D_Z  \rangle+h.c.
\eea

\medskip\noindent
where
$\langle D_Z\rangle \equiv \cos\theta_W\,\,\langle
D_2^3\rangle-\sin \theta_W\, \langle D_1\rangle
=-(m_Z^2/{g})\cos 2\beta+\cO(1/f)
$. The decay rate is then
\bea
\Gamma_{Z\ra \psi_X\psi_X}
&=&
\frac{m_Z}{24\, \pi\,g^2}
\bigg[\frac{m_Z^4}{2\,f^2}\bigg]^2
\cos^2 2\beta
\eea

\medskip\noindent
in agreement with previous results obtained for $B=0$
\cite{Luty:1998np,SK,Brignole:2003cm}.
The decay rate is independent of $m_A$ and should be within
the LEP error for $\Gamma_Z$
($2.3$ MeV \cite{pdg}).
One can then easily see that the increase of the Higgs
mass above the LEP bound (114.4 GeV) seen earlier in
Figure~\ref{higgs1} is consistent with the  current bounds for this
decay rate, which thus places only  mild constraints on $f$, below the
TeV scale ($\approx 200$ GeV) \cite{Luty:1998np,Brignole:2003cm}.

Similarly,
 $\cL^{new}$ can also induce Higgs decays into goldstino
pairs. The terms in $\cL^{new}$
that contribute to Higgs decays are $\cL_{F\,(2)}^{aux}$,
$\cL_D^{aux}$, $\cL_m^{extra}$ together with
the MSSM higgsino-Higgs-gaugino  coupling (last term in (\ref{tt})).
After using (\ref{Gcom}), expanding the Higgs
fields about their VEV, one finds:
\medskip
\bea\label{lcr}
\cL^{new}+\cL^{onshell}_0\supset
\frac{\mu\,v}{4\,f^2}\,m_A^2\,\cos
2\beta\,\,\opsi_X\opsi_X\,\Big[h_1^{0\,\prime}\sin\beta
-  h_2^{0\,\prime}\,\cos\beta\Big]
+h.c.+\cO(1/f^3)
\eea

\medskip\noindent
which, similarly to $Z$ couplings, is independent of gaugino masses.
Here $v=246$ GeV and
$h_i^0=1/\sqrt{2}\,\,(v_i+h_i^{0\, \prime}+i \sigma_i)$, $\langle
h_i^{0\,\prime}\rangle=0$,  $\langle \sigma_i\rangle=0$.
In the mass eigenstates basis one simply replaces
the square bracket in (\ref{lcr}) by $\big[
H^0\sin(\beta-\alpha)-h^0\,\cos(\beta-\alpha)\big]$.
One can also replace $m_A$ by
$m_A^2=m_h^2+m_H^2-m_Z^2+\cO(1/f^2)$,
where the Higgs masses can be taken to be  the MSSM values
(up to higher order corrections in $1/f$).
The decay rate of $h^0$ into a
pair of goldstinos is then
\bea
\Gamma_{h^0\ra
  \psi_X\psi_X}=\frac{m_h}{8\pi\,f^4}\,g_{h^0\psi_X\psi_X}^2
\eea

\medskip\noindent
where $g_{h^0\psi_X\psi_X}$ is the coupling of $h^0\psi_X\psi_X$ of
the above Lagrangian. For relevant values of $f$ above $\sim$1 TeV it turns
out that this decay rate is very small  relative to
other partial decay rates of the Higgs in the MSSM/SM. For example, for
a total decay rate near $10^{-3}$ GeV (valid near a Higgs
mass of order $\cO(100)$ GeV), the branching ratio of this decay mode is
well below  the usual ones and below that of SM Higgs going
 into $\gamma\gamma$, by a factor $\approx 10^{-3}-10^{-2}$.

\section{Conclusions}\label{conclusions}

In this work we performed a model independent analysis of the
consequences of a  light goldstino (of mass $\sim f/M_{Planck}$)
and investigated its couplings to the MSSM superfields.
This was done by treating $\sqrt f$ as a free parameter
that  can be as low as few times the soft SUSY breaking scale 
$m_{soft}\sim$ TeV. The formalism parametrized but did
not predict the soft masses, assumed to be fixed (near the TeV scale) by
an otherwise  arbitrary SUSY breaking sector.

The goldstino couplings can be determined by non-linear supersymmetry.
Above the $m_{soft}$ scale, one has the usual MSSM superfields
and the goldstino couples to them, while below this scale
the SM superpartners are integrated out and one is left with the
goldstino coupled to SM fields.
Both these cases can actually  be treated using constrained
superfields, where the constraints
effectively integrate out the corresponding superpartners
 in terms of light degrees of freedom.
For energy regimes $E\sim m_{soft}\leq \sqrt f$
the only constrained superfield is that of goldstino,
which couples to the MSSM superfields via the soft terms.
Below this energy regime additional constraints should be imposed on
the MSSM superfields themselves.
If supersymmetry breaking scale is low $\sqrt f\sim $ few TeV,
the goldstino couplings to the MSSM become important.
In this paper the leading couplings of all MSSM fields to the goldstino
were computed to $1/f^2$ order,  and these can be used for
phenomenological studies. In the limit  the hidden sector SUSY scale
is large with fixed soft masses,  the MSSM with explicit soft breaking
terms is recovered.

A significant impact of the aforementioned couplings  turned
out to be in the Higgs sector of the MSSM. It was noticed that the
usual MSSM scalar potential acquires additional terms involving
Higgs quartic couplings,  with coefficients depending on the ratio of the
soft masses to the `hidden' sector SUSY breaking scale ($\sqrt f$).
The presence of these couplings, effectively generated by integrating
out the sgoldstino via its superfield constraint,
can have  a significant impact  for the
Higgs mass and electroweak scale fine-tuning of the MSSM and these were
investigated in detail.

The masses  of the CP even and CP odd higgses were computed to order
$1/f^2$. It was shown that  for a low scale of SUSY breaking,
the SM-like Higgs mass $m_h$ is increased, to reach
and  cross the LEP bound, already at the tree level.
For values of $\sqrt f$ between 1.5~TeV~to~7~TeV one obtains a value of 
$m_h$ above the LEP bound. The correction increases with 
 $\mu$ and can remain significant even above this energy range.
As in the MSSM, quantum corrections increase $m_h$ further.
The quartic Higgs coupling was also increased by an additional (effective)
contribution related to the Higgs soft  terms. The benefit of this is that
the amount of fine  tuning of  the electroweak scale is then  reduced
relative to that in the MSSM alone, 
by a factor comparable to (or even larger)
than that due to the MSSM quantum corrections to the quartic Higgs coupling.
This can be easily understood if we recall that the main source of
fine tuning in the MSSM is related to the smallness of the MSSM Higgs
quartic coupling.

  The mechanism by which the mass of the SM-like
Higgs is increased and the fine-tuning reduced has similarities with
the method of additional effective operators usually considered in
the MSSM Higgs sector, to solve these problems.
 Indeed, using effective operators of dimension 
$d=5$ suppressed by  ``new physics'' at a scale $M$, one can increase the
Higgs mass above the LEP bound. 
The advantage in our case is that no new scale is introduced
in the visible sector.
In both cases the new scales introduced ($M$ or $\sqrt f$)
have comparable values, because in both cases the required increase for $m_h$
to be above the LEP bound is done via couplings that depend on the ratio 
$\mu/M$ and $\mu/\sqrt f$, respectively.

The possibility of an invisible decay of the MSSM lightest
Higgs or of $Z$ boson into a goldstino and the (lightest)
neutralino was investigated.
The decay rate of the Higgs can become comparable
to the $h^0\ra \gamma\gamma$ partial decay mode while that of the $Z$ boson
was shown to bring a lower bound on $\sqrt f\sim 700$ GeV, which is
stronger than previous similar bounds on $\sqrt f$. This bound is  consistent
with that required for a classical increase of $m_h$ above the LEP
bound, and does not apply in the case  $m_{\chi_1^0}>m_Z$ (if for
example $m_{\lambda_1}$ is large enough).
Higher order (in $1/f$) processes such as Z or Higgs decay into pairs
of goldstinos were also analyzed; these were  found to be too small to bring
constraints on $\sqrt f$ (for Z case) or sub-leading to the
decay into goldstino-neutralino (for the Higgs case), and in agreement
with previous calculations.

Let us mention that although we treated all MSSM superfields
in the linear SUSY realization ({\it i.e.} squarks and sleptons
lighter than $\sqrt{f}$), 
our results on Higgs mass and invisible
decays of the Higgs and $Z$ bosons
are largely  independent of this assumption.
Even if the quarks and leptons superfields are treated in the
nonlinear SUSY realization  ({\it i.e.}
squarks and slepton masses are large enough to be integrated out),
these results are not changed.

\newpage
\section*{Acknowledgments}

This work was supported in part by the European Commission under contracts
PITN-GA-2009-237920,  MRTN-CT-2006-035863 and ERC Advanced Grant
226371 (``MassTeV''), by INTAS grant 03-51-6346, by ANR (CNRS-USAR)
contract 05-BLAN-007901, by CNRS PICS no. 3747 and 4172.
During the last stages of this work, it was also partially supported
by the U.S. National Science Foundation under Grant No. PHY05-51164.\,
I.A. and E.D. would like to thank the KITP of UC Santa Barbara
for hospitality during  the last part of this work.
P.T. would like to thank the `Propondis' Foundation for its support.
D.G. thanks the CERN Theory Division for the financial support.
D.G. also thanks S.\,P. Martin, Z. Komargodski and  N.~Seiberg for
clarifications on goldstino couplings in the constrained superfield
formalism and their implications for phenomenology.

\section*{Appendix}

From the Higgs part of the scalar potential, eq.~(\ref{potential0}),
one finds the exact (in $1/f$) form of the  masses of the
CP even Higgs fields:
\medskip
\bea
m_{h,H}^2&=&
-2\mu^2\mp \frac{1}{4}\,\sqrt{\sigma}
+\,\frac{-2 B}{8\sin
  2\beta}\,\Big[3+\sqrt w_0 \,\Big]
+\frac{4\mu^4-B^2+2 \mu^2\,m_Z^2 +B\,m_Z^2 \sin 2\beta}{
4\,\mu^2+2\,m_Z^2\cos^2 2\beta+2 \,B\,\sin 2\beta}
\nonumber\\
&+& (1- \sqrt w_0)\,\, \frac{2\,f^2}{v^2}\,\,
\frac{
B^2+4 \mu^4
+ m_Z^2\,(4\mu^2 +m_Z^2)\,\cos^2 2\beta
+4 \,B\,\mu^2\sin 2\beta}{
\big(4\mu^2+  2\,m_Z^2\,\cos^2 2\beta+ 2\,B\,\sin 2\beta\big)^2}
\eea

\medskip\noindent
with the upper sign for the lightest $m_h$. The following notation was
used
\bea
w_0\equiv
1-\frac{ v^2}{f^2}\,\big( 4\,\mu^2+2\,m_Z^2\,\cos^2
2\beta+2\,B\,\sin 2\beta\big)
\eea
and
\bea
\sigma\!\!&=& \!\!\!
2 \,\Big[ 2\,x_1^2+8 B^2\!+5\,m_Z^4\!
+m_Z^2 \,\big(8 \,x_1\,\cos 2\beta+\!3 \,m_Z^2\,\cos 4\beta
-8 \,B\,\sin 2\beta\big)\Big]\!+\!\frac{v^4}{2\,f^4}
\Big[2\,x_2^4+5 B^4
\nonumber\\
&+&\!\! 20\,x_2^2\,x_1^2+5\,x_1^4
+
12\, x_1\,x_2\,(2 x_1^2
+2 B^2+x_2^2)\,\cos 2\beta
+ 3 (x_1^2-B^2)(x_1^2+B^2+2 x_2^2)\cos
4\beta
\nonumber\\
&+&\!\!
 10 B^2\,(x_1^2+2\,x_2^2)
+ 12 B\big(
2 (x_1^2+B^2)\,x_2+x_2^3
+x_1\,(x_1^2+B^2+2 x_2^2)\,\cos 2\beta
\big)\sin 2\beta\Big]
\nonumber\\[-4pt]
&+&
\frac{2 \,v^2}{f^2}
\Big[
-4 B^2\,(m_Z^2-3 x_2)+
m_Z^2\,x_2^2+x_1^2(5 m_Z^2+6 x_2)
+ 3 m_Z^2(x_1^2+x_2^2) \cos 4\beta
\nonumber\\
&+& 2 x_1 \big(2 x_1^2+5 B^2+
x_2 (6 m_Z^2 +x_2)\big)\cos 2\beta
+ 2 B\,\big(x_1^2+4 B^2
-3 m_Z^2\,x_2+2x_2^2
\nonumber\\
&+&3 x_1\,m_Z^2 \cos 2\beta\big)\sin
2\beta\Big],\qquad\quad\,\,\,\textrm{with}\qquad
x_1\equiv m_1^2-m_2^2,\quad\,x_2\equiv m_1^2+m_2^2.
\eea

\medskip\noindent
At the minimum of the scalar potential $x_{1,2}$ take
the values given in the text, Section~\ref{section4}.
A series expansion (in $1/f$) of $m^2_{h,H}$ was presented in the text.

\end{document}